\documentclass[pdflatex,sn-mathphys-num]{sn-jnl}

\usepackage{graphicx}%
\usepackage{multirow}%
\usepackage{amsmath,amssymb,amsfonts}%
\usepackage{amsthm}%
\usepackage{mathrsfs}%
\usepackage[title]{appendix}%
\usepackage{xcolor}%
\usepackage{textcomp}%
\usepackage{manyfoot}%
\usepackage{algorithm}%
\usepackage{algorithmicx}%
\usepackage{algpseudocode}%
\usepackage{listings}%
\usepackage{makecell}
\usepackage{comment}
\usepackage{array}
\usepackage{longtable}
\usepackage{comment}
\usepackage{multirow}
\usepackage{enumitem}
\usepackage{algorithm}
\usepackage{algpseudocode}
\usepackage[most]{tcolorbox}
\usepackage{listings}
\usepackage{float}
\usepackage{placeins}

\usepackage{tabularx}
\usepackage{booktabs} 
\received{Date Month Year}
\revised{Date Month Year}

\accepted{Date Month Year}

\definecolor{codebg}{RGB}{248,248,248}
\definecolor{codeframe}{RGB}{220,220,220}
\definecolor{codekeyword}{RGB}{0,0,120}
\definecolor{codecomment}{RGB}{90,90,90}
\definecolor{codestring}{RGB}{120,40,40}

\lstdefinelanguage{SPARQL}{
	morekeywords={
		PREFIX,SELECT,DISTINCT,WHERE,OPTIONAL,VALUES,FILTER,
		ORDER,BY,GROUP,LIMIT,OFFSET,ASK,CONSTRUCT,DESCRIBE
	},
	sensitive=true,
	morecomment=[l]{\#},
	morestring=[b]"
}

\lstdefinestyle{sparqlstyle}{
	language=SPARQL,
	backgroundcolor=\color{codebg},
	frame=single,
	rulecolor=\color{codeframe},
	basicstyle=\ttfamily\small,
	keywordstyle=\color{codekeyword}\bfseries,
	commentstyle=\color{codecomment}\itshape,
	stringstyle=\color{codestring},
	columns=fullflexible,
	keepspaces=true,
	breaklines=true,
	showstringspaces=false
}

\usepackage{tikz}
\usetikzlibrary{shapes.geometric, arrows, positioning}
\usepackage{tikz}
\usetikzlibrary{shapes.geometric, arrows, positioning}
\tikzstyle{layer} = [rectangle, rounded corners, minimum width=6cm, minimum height=1cm, text centered, draw=black, fill=blue!20]
\tikzstyle{human} = [rectangle, minimum width=3cm, minimum height=1cm, text centered, draw=black, fill=green!20]
\tikzstyle{ai} = [rectangle, minimum width=3cm, minimum height=1cm, text centered, draw=black, fill=orange!20]
\tikzstyle{arrow} = [thick,->,>=stealth]

\def\tsc#1{\csdef{#1}{\textsc{\lowercase{#1}}\xspace}}
\tsc{WGM}
\tsc{QE}
\tsc{EP}
\tsc{PMS}
\tsc{BEC}
\tsc{DE}

\theoremstyle{thmstyleone}%
\theoremstyle{thmstyletwo}%

\theoremstyle{thmstylethree}%
\newtheorem{definition}{Definition}%

\begin{document}

\title[From Prompts to Context: An Ontology-Driven Framework for Human-Generative AI Collaboration]{From Prompts to Context: An Ontology-Driven Framework for Human-Generative AI Collaboration}


\author*[1,2]{\fnm{Ng\d{o}c Luy\d{\^{e}}n} \sur{L\^e}}\email{ngoc-luyen.le@hds.utc.fr}

\author[2]{\fnm{Marie-H\'el\`ene} \sur{Abel}}\email{marie-helene.abel@hds.utc.fr}

\author[1,3]{\fnm{Bertrand} \sur{Laforge}}\email{laforge@lpnhe.in2p3.fr}

\affil[1]{Gamaizer, 93340 Le Raincy, France}

\affil[2]{Universit\'{e} de technologie de Compi\`egne, CNRS, Heudiasyc (Heuristics and Diagnosis of Complex Systems), CS 60319 - 60203 Compi\`egne Cedex, France}

\affil[3]{Sorbonne Universit\'{e}, CNRS UMR 7585, LPMHE (Laboratoire de Physique Nucl\'{e}aire et des Hautes \'{E}nergies), 75252 Paris cedex 05, France}


\abstract{Collaborations with Generative AI often begin with a short prompt and end with an opaque output, leaving implicit who was involved, what task was being pursued, which resources were used, and which constraints should have shaped the process. This limited contextual explicitness hinders trust, traceability, and accountability, particularly when Generative AI is embedded in information-intensive workflows such as search, querying, and profile management. This paper introduces \emph{From Prompts to Context}, an ontology-driven framework for representing Human--Generative AI collaboration. Its core component, the Contextual Collaboration AI Ontology (CCAI), models key elements of collaboration -- including tasks, agent roles, resources, and constraints -- as a shared machine-interpretable vocabulary. By combining populated CCAI instances with SPARQL-based context retrieval in operational workflows, the framework turns otherwise ephemeral prompt-response interactions into structured and queryable collaboration traces linking prompts, outputs, and their surrounding context. The approach is illustrated through a case study involving a software development team building a competency-based education feature for viewing and updating learner competency profiles. The case study shows how the framework can support the representation and documentation of collaboration episodes across requirements analysis, design, implementation, and testing. Within this illustrative setting, the results indicate that explicit collaboration modelling helps make task context more explicit, improves the traceability of AI-generated contributions, and supports more transparent and accountable Human--Generative AI practices. We conclude by outlining design principles for future Human--Generative AI systems that emphasise not only output quality, but also the explicit representation of the collaborative context in which outputs are produced.}




\maketitle

\section{Introduction}

Over the past few years, advances in artificial intelligence have shifted from narrow rule-based systems to foundation models capable of generating human-like text, programming code, images, and other creative outputs \cite{bommasani2021opportunities}. These Generative AI models -- typified by large language models \cite{brown2020language} and Transformer-based architectures \cite{vaswani2017attention} -- offer new opportunities for content creation, problem-solving, and decision support \cite{thirunavukarasu2023large,hu2024agentscomerge,yao2024tree}. However, as these models become increasingly integrated into workflows, the interactions between human collaborators and AI-enabled systems have grown more complex, raising questions about how to supervise AI outputs, enforce constraints, and delineate roles and responsibilities \cite{ozmen2023six}. A recurring problem is that collaborations often begin with a prompt and end with an opaque output, leaving little evidence of how a result was produced or why a recommendation should be trusted, and providing no explicit representation of the collaboration context in which the AI was used.

In this paper, we distinguish between generative AI models and generative AI systems. Generative AI models refer to the underlying neural models that produce outputs such as text, code, images, or other content from prompts. By contrast, generative AI systems denote broader software or socio-technical systems in which one or more generative models may be embedded together with additional components such as retrieval mechanisms, external tools, memory, or iterative workflows. Some of these systems may be agentic when they involve planning, tool use, or multi-step autonomous execution. This distinction is important because the opacity observed in Human--Generative AI collaboration often arises not only from the model itself, but from the broader system-level interaction among prompts, retrieved context, intermediate processing steps, tool calls, and human interventions \cite{sapkota2025ai}.

In many real-world applications, humans rely on Generative AI for drafting, ideation, or augmenting specific tasks. For instance, a product team may use Generative AI to suggest user-facing text for reports or documentation \cite{holmstrom2024organizations}; an educator may employ Generative AI to generate tailored lesson plans for students \cite{elsayary2024investigation}; and a programmer may collaborate with Generative AI to write code \cite{dong2024self}. In these scenarios, a shared framework that explicitly models roles, tasks, feedback loops, and ethical considerations is essential to ensure that human oversight remains central, that suggestions are transparent, and that collaboration goals are clearly defined, especially when these activities are embedded in information-centric workflows such as querying data, updating profiles, or generating explanations.

In the context of Human--Generative AI collaboration, ontologies can serve as structured frameworks that formally define the concepts, roles, and relationships governing each stage of the interaction \cite{gruber1993translation}. By establishing a shared vocabulary and clear structures, ontologies help delineate human responsibilities and AI capabilities, ensuring both parties have a common reference for tasks, feedback loops, and domain-specific requirements. This supports accountable practice, as constraints and decision-making processes can be explicitly encoded \cite{wu2022ai}. In particular, making collaboration context first-class -- so that artifacts are linked to tasks, roles, prompts, sources, and responsible agents -- enables traceability and auditable review.

Despite growing interest in Human--AI collaboration \cite{cabrera2023improving,vaccaro2024combinations,mozannar2024effective}, current approaches often lack a standardized method for capturing and communicating the contextual elements that govern how people and AI systems work together. This gap can lead to ad hoc solutions where essential aspects -- such as ethical guidelines, version control, or domain assumptions -- are lost or inconsistently tracked over the course of a project \cite{buschek2021nine,kulkarni2023word}. As a result, organizations struggle to maintain traceability and accountability, while researchers find it difficult to compare results and replicate studies across collaborative environments.

To address these issues, this paper introduces ``\textit{From Prompts to Context: An Ontology-Driven Framework for Human--Generative AI Collaboration}'', which develops a foundational ontology that adapts to varied collaborative contexts. The Contextual Collaboration AI Ontology (CCAI) captures the who, what, how, and why of each stage in the collaboration lifecycle. In our approach, the ontology is used not only to represent collaboration context, but also to retrieve task-relevant semantic information that can be injected into prompts, thereby grounding AI outputs in explicit roles, resources, and constraints. By leveraging semantic web standards and SPARQL to retrieve context, and by introducing classes for tasks, feedback loops, AI-generated outputs, and constraints, the framework integrates ontology-derived information into day-to-day work so that prompts and outputs are semantically grounded and explicitly linked to their surrounding collaboration episodes \cite{988453,chen2024large}. Interface elements (a query box, an explainer panel, and a prompt template) are used as instrumentation to expose context and collect auditable traces, rather than as a full user interface contribution.

To ground our framework in a realistic context, we present a case study of a software team building a competency-based education management feature ``\textit{View \& Update Competency Profiles}''. We use this case to demonstrate and analyze the application of our ontology-driven workflow. We analyze collaboration artifacts (prompts, AI-generated tests, commits, documentation) and ontology instances, and we use targeted SPARQL queries to recover operational context and reconstruct decision trails. Our findings indicate that, in this illustrative case study, the framework helped make task context more explicit, supported the traceability of AI-generated contributions, and enabled more transparent reviews .Our findings indicate that, in this illustrative case study, the framework helped reduce contextual ambiguity, improve the traceability of AI-generated contributions, and support more transparent reviews?outcomes aligned with the framework?s goals. We intentionally do not claim efficiency or productivity gains; rather, the study focuses on transparency, traceability, and shared understanding within a bounded setting.

This paper makes three contributions: (1) a contextual ontology and framework that make collaboration semantics and context first-class for Human--Generative AI work; (2) an instrumentation approach that integrates SPARQL-derived context into prompting and review to produce structured, auditable collaboration traces; and (3) an in-depth case study that demonstrates the application of our framework, highlighting its potential to enhance transparency and accountability in practice and offering design implications for future systems.

In the following sections, we discuss the theoretical underpinnings of Human--AI collaboration and review relevant ontologies and related work. We then introduce our methodology for a Human--Generative AI collaboration framework, focusing on how contextual factors are modelled and integrated to support dynamic, transparent, and ethically grounded interactions. Next, we present a representative use case, demonstrating how the proposed ontology-based framework enhances transparency, iterative feedback, and ethical oversight. Finally, we conclude by summarizing the implications of our findings and outlining potential avenues for future research and refinement.

\section{Related Work}
Generative AI systems like ChatGPT, Claude AI, and Gemini have shown remarkable capabilities in areas such as content creation, decision support, and collaborative problem-solving~\cite{holmstrom2024organizations,elsayary2024investigation}. 
This section presents the key developments in Generative AI, Human-Generative AI collaboration, focusing on the role of ontologies and contextual awareness in enhancing collaborative efficiency and trustworthiness.

\subsection{Generative AI}
Generative AI generally refers to computational models or systems capable of generating new content -- such as human-like text, images, audio, video, or programming code -- by learning patterns and structures from very large training data \cite{gozalo2023chatgpt}. Significant advancements in this field have emerged from various deep learning architectures, including diffusion models \cite{ho2020denoising}, variational autoencoders \cite{kingma2019introduction}, and generative adversarial networks \cite{goodfellow2020generative} for image and video inputs. Moreover, Transformer-based architectures -- such as GPT (Generative Pre-Trained Transformer) \cite{radford2018improving} and BERT \cite{devlin2018bert} -- have played a pivotal role in natural language processing tasks, providing the foundation for large language models that generate and understand text-based content.

Contemporary generative models, especially large language models (LLMs), operate as probabilistic autoregressive systems that generate outputs token by token from a prompt and its preceding context. Their capabilities are typically acquired through large-scale pre-training on heterogeneous corpora and are often later adjusted through instruction tuning or other forms of human-feedback-based alignment. These same design properties help explain several well-known limitations. Because outputs are generated from learned statistical regularities rather than grounded world models, LLMs may produce fluent but factually unsupported content, commonly described as hallucinations. They may also reproduce biases present in their training data, while alignment procedures based on human feedback can introduce additional effects such as sycophancy \cite{sharma2023towards} or imperfect reflection of diverse human viewpoints \cite{santurkar2023whose}. In collaborative settings, these limitations motivate the need for external mechanisms that make tasks, resources, roles, and constraints explicit, so as to reduce prompt under-specification and improve the contextual grounding of generated outputs.

\FloatBarrier
\begin{table}[!h]
\caption{Categories of Generative AI by Input and Output.}
	\label{tab:genai_io}
	\begin{tabular*}{\textwidth}{@{}p{1.2cm}p{1.2cm}p{3.0cm}p{3.0cm}p{2.7cm}@{}}
		\toprule
		\textbf{Input}              & \textbf{Output}          & \textbf{Description}                                           & \textbf{Applications}                                         & \textbf{Examples}                \\ \midrule
		Text                        & Text                     & Generates or transforms text based on prompt inputs.               & Chatbots, content creation, text summarization, language translation.             & ChatGPT~\cite{achiam2023gpt}, Gemini~\cite{team2023gemini}, Mistral~\cite{jiang2023mistral}                        \\ \midrule
		Text                        & Image/ Video              & Creates visual content from descriptive prompts.               & Graphic design, advertising, media production.                & DALL-E~\cite{betker2023improving}, Sora~\cite{brooks2024video}                                 \\ \midrule
		Image/ Video                 & Text                     & Interprets visual inputs to generate textual descriptions.      & Accessibility tools, multimedia analysis.                     & ChatGPT~\cite{achiam2023gpt}, Gemini~\cite{team2023gemini}                  \\ \midrule
		Text                        & Audio                    & Converts text into synthesized speech or music.                & Virtual assistants, personalized learning tools.               & Tacotron~\cite{wang2017tacotron}, AudioGen~\cite{kreuk2022audiogen}                     \\ \midrule
		Audio                       & Text                     & Transcribes spoken language or sounds into text.               & Automated transcription, accessibility for hearing-impaired.   & Whisper~\cite{radford2023robust}, LauraGPT~\cite{du2023lauragpt}                              \\ \midrule
		Audio                       & Audio                    & Enhances or transforms audio signals.                          & Noise reduction, voice modulation, sound effects.              & WaveNet~\cite{van2016wavenet}, MusicGen~\cite{copet2024simple}                             \\\midrule
		Image                       & Image                    & Transforms or enhances input images into new formats.           & Style transfer, image restoration, medical imaging.            & Latent Diffusion~\cite{rombach2022high}, Style-based GAN2~\cite{karras2019style}               \\ \midrule
		Text/ Image                  & Video                    & Produces video from descriptive prompts or images.             & Video storytelling, gaming, advertising.                      & Imagen Video~\cite{ho2022imagen}, Make-A-Video~\cite{singer2022make}                   \\ \midrule
		Text                        & 3D Models                & Generates 3D models from textual descriptions.                 & Gaming, virtual reality, product design.                       & DreamFusion~\cite{poole2022dreamfusion}, Point-E~\cite{nichol2022point}                         \\ \midrule
		Text                        & Program-ming Code         & Writes or completes programming code from text prompts.        & Coding assistance, automating repetitive tasks.                & Github Copilot, CodeT5~\cite{wang2023codet5}                      \\ \midrule
		Multi- modal & Multi- modal Output        & Combines multiple inputs to generate enriched outputs.         & Educational tools, intelligent virtual assistants.             & ChatGPT~\cite{achiam2023gpt}, Flamingo~\cite{alayrac2022flamingo}                               \\ 
		\bottomrule
	\end{tabular*}
\end{table}
\FloatBarrier

Generative AI models and systems are designed to produce diverse content by transforming specific inputs and contextual prompts into meaningful outputs. The relationship between inputs and outputs varies across applications, from simple one-to-one mappings to more complex, dynamic interactions \cite{bandi2023power}. Table \ref{tab:genai_io} showcases various input-output configurations for different types of Generative AI. For instance, in text-based applications, these systems can create coherent paragraphs, summaries, or creative stories from a single sentence or a few keywords, commonly referred to as prompts. Their flexibility allows them to adapt to different input types and leverage learned patterns to generate diverse, high-quality outputs, driving innovation and efficiency across many domains.

Despite their impressive capabilities, Generative AI systems face several potential pitfalls that can impact their effectiveness and reliability: (i) Bias is a significant concern, as these models often inherit biases present in their training data, leading to outputs that may reinforce stereotypes or provide unequal treatment across demographic groups \cite{gorska2025ai}; (ii) Lack of context can result in outputs that are irrelevant or inappropriate for specific tasks, as the models struggle to fully understand nuanced or domain-specific requirements \cite{bender2021dangers}; and (iii) Hallucination, a phenomenon where the AI generates plausible-sounding but factually incorrect or completely fabricated information \cite{xu2024hallucination}. This issue can be particularly problematic in critical applications such as healthcare, legal, or financial contexts, where accuracy and trustworthiness are paramount.

Addressing these challenges requires careful dataset curation, robust fine-tuning, transparent system design, and ongoing human supervision to ensure adherence to ethical constraints and support the practical use of Generative AI systems. By mitigating and overcoming these shortcomings, Generative AI can collaborate effectively with humans in shared environments to build, implement, and optimize workflows, tasks, and projects. In the next section, we will explore the existing literature on Human and Generative AI collaboration.

\subsection{Human-Generative AI Collaboration}

Generative AI, a relatively recent advancement in artificial intelligence, has gained significant traction in recent years. Despite its relatively short history, collaboration between humans and Generative AI has already been applied across numerous domains, investigating how humans and generative models can work together to accomplish tasks, support decision-making, and solve complex problems \cite{reverberi2022experimental,thirunavukarasu2023large,hu2024agentscomerge, yao2024tree}.
This emerging field draws upon decades of foundational research in Human-Computer Interaction (HCI), Computer-Supported Cooperative Work (CSCW), and Artificial Intelligence (AI). These earlier studies laid the groundwork for understanding how technology could enhance human capabilities, providing a theoretical and practical basis for today's Human-Generative AI collaboration efforts \cite{song2024human,cila2022designing}.

Recent research has explored how Humans and Generative AI systems can collaborate effectively across diverse fields, providing valuable insights into the dynamics of these interactions. Studies in creative domains such as comprehension and creative writing \cite{chang2021exploring,yang2022ai,li2024value}, audio production \cite{suh2021ai}, video generation \cite{singer2022make}, software development \cite{Kim11072024,FRANCE2024649}, and problem-solving \cite{xu2024chatglmmath} highlight both the benefits and challenges of Human-Generative AI collaboration. 
These benefits may include support for ideation, increased productivity, and the ability to explore novel solutions. However, recent studies suggest that while generative AI can improve individual outputs in some settings, it may also reduce the overall diversity of ideas and lead to more homogeneous patterns across outputs \cite{doshi2024generative,meincke2025chatgpt}. However, challenges such as trust calibration, effective workflow integration, and the development of accurate mental models for Generative AI capabilities remain significant obstacles \cite{10.1145/3290605.3300233,10.5555/3061053.3061219}. Addressing these challenges requires carefully designed interfaces, processes and interaction frameworks that foster intuitive and efficient collaboration.

A central focus of this work is understanding the respective roles of humans and Generative AI in collaborative scenarios. Generative AI excels in tasks that require pattern recognition, data analysis, and the generation of diverse outputs, whereas humans contribute essential strengths in judgment, contextual understanding, and creative direction \cite{Shneiderman_2020}. Effective collaboration depends on leveraging these complementary capabilities while ensuring human agency and maintaining meaningful control over decision-making processes \cite{10.1145/3432945}. Achieving this balance can enable seamless and productive Human-Generative AI partnerships across a wide range of applications.

An emerging strategy for enhancing collaboration is the integration of context-aware systems through ontologies. Ontologies provide a formal structure for modeling workflows, capturing tasks, roles, and goals, and creating a shared understanding between humans and Generative AI~\cite{zhong2024ontology}. This approach bridges the gap between human expertise and Generative AI capabilities, making collaboration more structured, efficient, and aligned with human values. In the next section, we delve into related work on how ontologies support collaborative environments.

\subsection{Ontologies and Collaborative Environments}
Ontologies have long been foundational in the fields of symbolic AI and knowledge representation, where they are used to create taxonomies and structured vocabularies for organizing information, enabling inference, and reasoning over data \cite{gruber1993translation, STUDER1998161}. Their ability to formally define entities, relationships, and processes within a specific domain makes them a powerful tool for structuring and supporting collaborative environments. By providing a shared conceptualization that is both human- and machine-readable, ontologies enable interoperability and a common understanding among diverse agents, whether human or artificial. This coordination can range from purely human-human collaboration (e.g., group decision-making or team-based workflows) to Human-AI collaboration (e.g., context-aware systems or collaborative robotics).  The following works illustrate how various researchers have approached the development and application of ontology-driven frameworks to enhance collaboration.

\renewcommand{\arraystretch}{1.2}
\begin{table}[h]
	\centering
	\caption{Comparison of Different Approaches to Developing Collaboration Ontologies.}
	\label{tab:collab_ontologies}
	\begin{tabular}{p{1.0cm} p{1.8cm} p{2.2cm} p{3.0cm} p{3.0cm}}
		\toprule
		\textbf{Works} 
		& \textbf{Domain/ Focus} 
		& \textbf{Collaboration Supported} 
		& \textbf{Key Features} 
		& \textbf{Limitations} \\
		\midrule
		
		Konate et al.~\cite{konate2020ontology} 
		& 
		Medical and general decision-making
		& 
		Primarily Human-Human; potential for AI-assisted workflows 
		& Structured representation of decision processes; emphasizes joint decision rationale
		& Limited explicit support for Generative AI
		\\\midrule
		
		Li et al.~\cite{li2022collaboration}
		& 
		Collaboration in Industry 4.0 Production Lines

		& 
		Human-related Collaboration
		& Defines context factors influencing collaboration, focus on environment and user roles
		& Does not deeply address Generative AI; Limited empirical validation
		\\\midrule
		
		Gu et al.~\cite{gu2020ontology}
		& 
		Intelligent / smart environments
		& 
		Human-AI (adaptive systems); potential extension to multi-agent
		& Focus on user profiles and personalization; context-driven adaptation
		&  Collaboration is implicit (collaborative environment-level)
		\\\midrule
		
		Oliveira et al.~\cite{oliveira2007towards}
		& 
		General collaboration frameworks
		& 
		Human-Human; foundational approach that could extend to AI
		& Basic concepts for group tasks, roles, and goals; Emphasis on knowledge sharing
		
		&  Needs refinement for Generative AI contexts;
		Primarily conceptual with limited case studies
		\\\midrule
		
		Gasmi et al.~\cite{gasmi2017ontology}
		& 
		Education and industry partnerships
		& 
		Human-Human, potential to add AI for data-driven insights
		& Models academic-industry collaboration activities; Lifecycle approach for projects
		
		& Limited mention of advanced AI collaboration; Domain-specific; less generalizable
		\\\midrule
		
		Olivares-Alarcos et al.~\cite{olivares2022ocra}
		& 
		Collaborative robotics (Human-Robot)
		& 
		Human-AI with physical robots (potential for Generative AI in control)
		& Defines roles, tasks, context for robot-human teams; Adaptive features for real-time changes
		& Focus on robotic domain; Generative AI aspects not explicitly covered
		\\\midrule
		Knoll et al.~\cite{knoll2010collaboration}
		& 
		Generic group support systems
		& 
		Primarily Human-Human; potential integration with AI
		& Framework for group interactions and decision support; Reusable knowledge-based collaboration patterns
		& 
		Does not explicitly address Generative AI; Requires adaptation for complex AI tasks
		\\
		\bottomrule
	\end{tabular}
\end{table}

Different works of collaboration ontologies, as summarized in Table~\ref{tab:collab_ontologies}, highlights the diverse approaches taken to model and support collaborative environments. Most ontologies primarily focus on human-human collaboration, offering structured representations of roles, tasks, and contexts to facilitate group tasks and decision-making. For instance, works like Konate et al.~\cite{konate2020ontology} and Oliveira et al.~\cite{oliveira2007towards} emphasize shared decision-making and knowledge-sharing frameworks that cater primarily to human interactions. However, these foundational ontologies often lack the      level of detail   needed to address Generative AI or dynamic, adaptive scenarios.

Some works, such as Gu et al.~\cite{gu2020ontology} and Olivares-Alarcos et al.~\cite{olivares2022ocra}, expand the scope to include Human-AI or Human-Robot collaboration, introducing adaptive features and context-awareness. These ontologies demonstrate potential for multi-agent systems and intelligent environments, where context-driven personalization is essential. For example, Gu et al.~\cite{gu2020ontology} focuses on user profiles and environment adaptation, aligning with the growing need for dynamic collaboration systems. Similarly, Olivares-Alarcos et al.~\cite{olivares2022ocra} explicitly addresses human-robot collaboration, defining roles and tasks within robot-human teams while incorporating real-time adaptability.

Despite these advancements, explicit support for Generative AI remains underexplored. While some frameworks hint at AI-assisted workflows, such as Li et al.~\cite{li2022collaboration} in Industry 4.0 and Gasmi et al.~\cite{gasmi2017ontology} in education-industry partnerships, they lack detailed integration of Generative AI capabilities, such as ideation or dynamic content generation. This limitation highlights a significant research gap, particularly given the increasing prominence of Generative AI in collaborative environments.


In summary, current research on collaboration ontologies reveals gaps, particularly the need to adapt ontologies for Generative AI by incorporating dynamic and generative capabilities to support creative and intelligent systems. Additionally, enhancing context-awareness within ontological frameworks is essential to effectively handle diverse and evolving scenarios. To address these limitations, we propose an ontology-driven framework in which tasks, resources, roles, and constraints are explicitly represented and queried to support context-aware prompt construction in Human--Generative AI collaboration. In the next section, we present a methodology for designing a context-aware collaboration ontology that integrates Generative AI capabilities.

\section{Methodology}
We present an approach to supporting Human-Generative AI collaboration. First, we introduce a Collaboration Framework that defines the key components and processes enabling effective interaction and shared workflows between humans and Generative AI. Next, we propose a Contextual Collaboration AI Ontology designed to structure and model the knowledge, roles, and processes involved in collaboration. Together, these elements provide a cohesive foundation for enhancing collaboration, ensuring clarity, adaptability, and efficiency.

\subsection{Human-Generative AI Collaboration Framework}
This section focuses on defining collaboration and exploring its various innovative aspects in the context of Human-Generative AI interaction. We then introduce the different levels of collaboration, followed by a general framework that encapsulates these interactions and provides a structured approach to facilitating effective Human-Generative AI partnerships.
\subsubsection{Human-Generative AI Collaboration: Definitions and Levels}

We begin by establishing a clear definition of collaboration within the unique context of Human-Generative AI interaction, focusing on its multifaceted nature and diverse levels of engagement. Recent work on co-creative AI also shows that Human--AI collaboration may involve multiple roles and interaction patterns rather than a single form of assistance. In particular, Lin and Riedl \cite{lin2023ontology} highlight that AI systems can participate in different capacities depending on how responsibilities and initiative are distributed between human and AI participants. This perspective helps refine the collaboration strategies considered in our framework. These include collaborative creation, where humans and Generative AI generate novel outputs together; Generative AI-powered augmentation, where Generative AI enhances human capabilities with insights and options; and iterative feedback, where humans refine and guide Generative AI outputs to achieve desired results~\cite{wang2020human,pangavhane2024ai}.

\begin{definition}
	Collaboration is the process of two or more individuals, teams, or organizations working together to achieve a common goal, share knowledge, or solve a problem. It involves the exchange of ideas, resources, and efforts, often leveraging diverse skills and perspectives to accomplish tasks more efficiently and effectively than working independently. Effective collaboration requires communication, trust, mutual respect, and a shared commitment to the goal~\cite{mattessich2018collaboration}.
\end{definition}

\begin{definition}
	Human-Generative AI Collaboration refers to the process of humans and generative artificial intelligence systems working together to achieve shared goals, solve problems, or create outputs~\cite{paulus2012collaborative,haase2024human}. This collaboration leverages the complementary strengths of humans -- such as creativity, critical thinking, and emotional intelligence  -- and Generative AI systems, which excel in processing large amounts of data, generating content, and performing repetitive or computationally intensive tasks.
\end{definition}

Human-Generative AI collaboration involves three key levels of interaction that maximize the strengths of both humans and Generative AI: (i) Co-creation, where humans and Generative AI jointly produce content, designs, or ideas, with humans providing guidance and refinement while Generative AI generates drafts or suggestions (e.g., an author brainstorming storylines with Generative AI or a designer refining Generative AI-generated concepts)~\cite{hosanagar2024designing}; (ii) Augmentation, where Generative AI enhances human capabilities by offering insights or options to support decision-making, allowing humans to focus on complex tasks (e.g., a financial analyst using Generative AI to identify market trends or a teacher personalizing learning paths with AI recommendations)~\cite{zhu2024human}; and (iii) Iterative Feedback, where humans refine Generative AI outputs through input and corrections to ensure alignment with specific needs (e.g., a marketer editing Generative AI-generated posts to match brand tone or a developer refining Generative AI-suggested code). These interactions, widely adopted in domains like creative industries, education, research, and decision-making, foster innovation and efficiency while maintaining human oversight. In the next section, we explore different levels of collaboration involving Generative AI~\cite{treude2025developers}.
\subsubsection{Generative AI Collaboration Levels}

\begin{figure}[h]
	\centering	
	\includegraphics[width=0.9\textwidth]{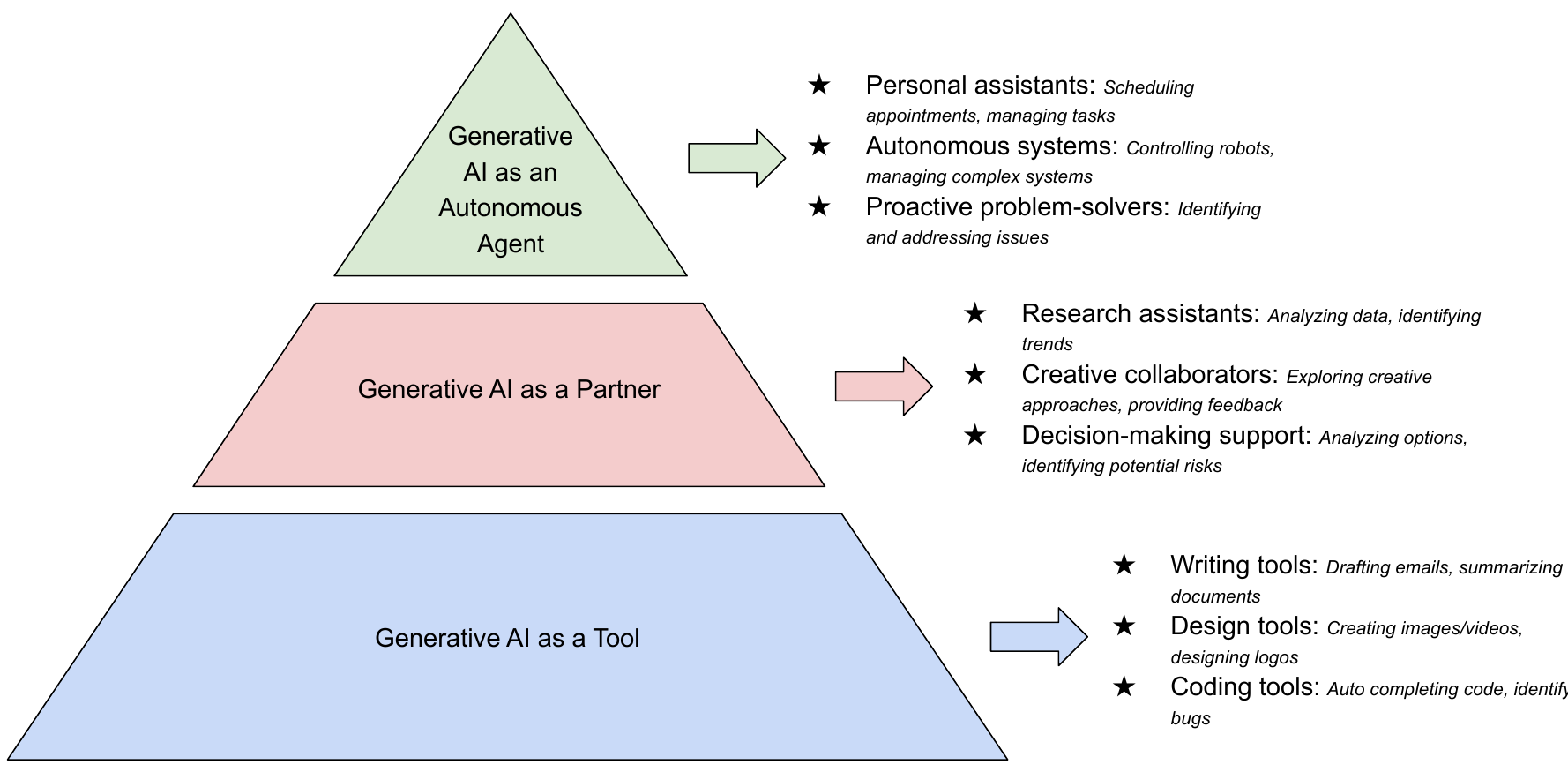}
	
	\caption{Different collaboration levels of Generative AI.}
	\label{fig_01}
\end{figure}

To better distinguish these forms of collaboration, we characterize them according to three criteria: (i) the degree of human specification of the task, (ii) the degree of initiative exercised by the AI during task execution, and (iii) the extent to which the AI contributes to shaping intermediate decisions rather than only executing bounded subtasks. Based on these criteria, we distinguish three collaboration levels: Generative AI as a Tool~\cite{jo2023promise}, Generative AI as a Partner~\cite{kilde2024generative}, and Generative AI as an Autonomous Agent~\cite{cronin2024autonomous}, as illustrated in Figure~\ref{fig_01}. These levels reflect increasing degrees of AI involvement, autonomy, and influence on the collaborative process.


At the Generative AI as a Tool level, the AI operates within tightly specified human instructions and supports bounded subtasks without substantially shaping the task itself. Its role is primarily assistive: it accelerates execution, automates repetitive operations, and provides localized suggestions under close human supervision. Typical examples include code autocompletion, grammar correction, prompt-based summarization, and interface mockup generation from clearly defined requirements. In this mode, humans retain control over both the objective and the operational boundaries of the task, while the AI contributes mainly through targeted assistance rather than broader co-development or decision-making. This level therefore reflects a bounded form of collaboration in which AI augments human work without actively participating in the broader framing of the problem or solution~\cite{sengar2024generative}.

Generative AI as a Partner represents a more interactive level of collaboration in which the AI contributes not only to task execution but also to exploration and iterative refinement within human-supervised boundaries~\cite{kilde2024generative}. In this role, Generative AI goes beyond bounded assistance by generating alternative drafts, proposing solution paths, highlighting trade-offs, and supporting ideation in ways that influence intermediate decisions. Humans, however, remain responsible for framing the objective, interpreting and refining the outputs, and making final decisions regarding their acceptance and use. Typical examples include research assistants that analyze data and suggest directions for further inquiry, creative collaborators that help explore alternative concepts, and decision-support systems that identify options, risks, or trade-offs. The key distinction from the Tool level is therefore not merely that outputs are produced, but that the AI plays a more active role in shaping the evolving problem-solving or creative process while still operating under human guidance.


At the Generative AI as an Autonomous Agent level, humans provide mainly high-level goals, constraints, or success criteria, while the AI plans, coordinates, and executes multiple steps with comparatively limited intervention~\cite{jabbour2024generative,cronin2024autonomous}. This level is distinguished from the previous two by the broader delegation of initiative and task coordination. In this role, the AI may autonomously generate outputs, select intermediate actions, and adapt its behavior across successive stages of a workflow, while humans remain responsible for oversight, validation, and the handling of ethical or organizational implications. Typical examples include systems that manage multi-step content generation pipelines, coordinate scheduling or monitoring tasks, support autonomous decision workflows, or operate in complex environments requiring continuous adaptation. The defining characteristic of this level is therefore not only increased automation, but the AI's capacity to organize and pursue task execution under high-level human guidance rather than detailed step-by-step instruction.


These levels of collaboration demonstrate the adaptability of Generative AI in meeting diverse human needs, ranging from enhancing individual productivity to enabling large-scale automation and fostering innovation. Together, they form the foundation for the proposed collaboration framework, which integrates these interaction levels into a cohesive model for Human-Generative AI partnerships, as detailed in the following section.

\subsubsection{Human-Generative AI Collaboration Framework}

The emergence of Generative AI has revolutionized collaboration by enabling novel and impactful interactions between humans and AI systems~\cite{fragiadakis2024evaluating}. In this section, we propose a Human-Generative AI Collaboration Framework, presenting a structured and dynamic model designed for seamless interaction. The framework leverages the unique strengths of both human agents and Generative AI agents to achieve shared objectives. At its core is a dynamic collaboration process, supported by functional layers that integrate human inputs, Generative AI capabilities, and shared resources. These layers aim to facilitate efficient task execution, informed decision-making, and iterative improvement, ensuring adaptive and effective collaboration.

At the center of the framework, the Collaboration Process Management serves as a central connecting layer that integrates various components, including tasks, resources, context management, decision-making support, and evaluation and feedback. This process is enriched by the complementary interplay between human agents and Generative AI agents, each contributing distinct but synergistic roles. Human agents bring critical contextual understanding, creativity, and oversight, ensuring the collaboration aligns with strategic goals and ethical principles. In contrast, Generative AI agents contribute scalability, speed, and innovative capabilities, generating diverse outputs and adapting dynamically to evolving requirements through iterative refinement.

\begin{figure}[h]
	\centering	
	\includegraphics[width=0.98\textwidth]{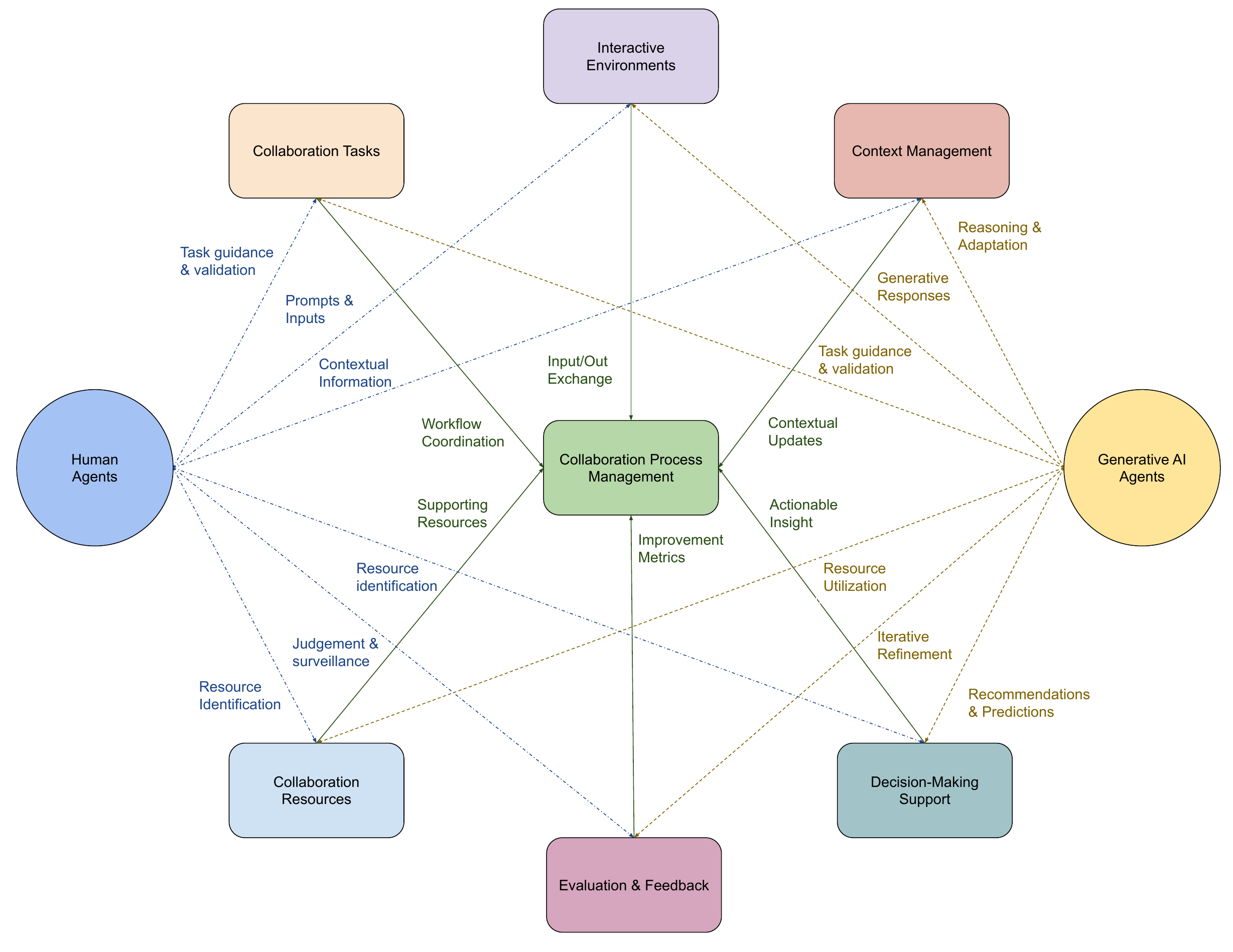}
	
	\caption{Proposed Human-Generative AI Collaboration Framework (Dash-dotted blue arrows represent human contributions; Dashed orange arrows: indicate Generative AI roles such as content generation, reasoning, and task execution, complementing human expertise with efficiency and creativity;
		Green arrows: depict the flow of outputs from functional layers to the collaboration process;)}
	\label{fig_00}
\end{figure}

The framework includes different functional layers, as illustrated in Figure~\ref{fig_00}, each playing an important role in the collaboration process: 
\begin{itemize}
	\item \textit{The Interactive Environment} serves as the primary interface where humans and Generative AI exchange inputs and outputs. It facilitates clear communication, allowing humans to provide prompts and feedback while receiving Generative AI-generated suggestions or solutions. 
	\item \textit{The Context Management layer} ensures that the collaboration adapts to the task's evolving context, incorporating environmental changes, collaboration goals, and real-time data. This dynamic adaptation is crucial for maintaining relevance and effectiveness in complex scenarios.
	\item \textit{The Collaboration Tasks layer} focuses on distributing roles, orchestrating workflows, and automating repetitive processes. Here, human agents validate AI-generated outputs and provide guidance, while Generative AI agents execute predefined tasks or generate creative alternatives.
	\item  \textit{The Decision Support layer} enables actionable insights and predictions to guide human decision-making, ensuring that the collaboration remains aligned with broader objectives.
	\item \textit{The Evaluation and Feedback layer} incorporates continuous monitoring and iterative improvement, enabling both agents to refine their contributions and enhance overall outcomes.
	\item \textit{The Collaboration Resources layer} provides the tools, datasets, and ontologies necessary for effective collaboration. These resources act as enablers, empowering human agents to contribute their expertise and allowing Generative AI agents to leverage structured knowledge for collaboration tasks.
\end{itemize}

Together, these layers form an interconnected system that enables dynamic, adaptive, and context-aware collaboration by integrating Human and Generative AI capabilities. The dynamic nature of the proposed framework allows it to respond in real-time to evolving tasks, user inputs, and environmental changes, ensuring relevance and flexibility. Its adaptive design enables continuous learning and improvement through iterative feedback loops, refining both processes and outputs to align with collaboration goals. Additionally, the context-aware functionality ensures that all components operate with a unified understanding of the collaboration task's context, including goals, resources, and constraints.

The interaction between human agents and Generative AI is crucial to the success of the framework. Humans offer creativity, judgment, and domain-specific expertise, while Generative AI provides computational power, diverse outputs, and the ability to scale tasks. The framework ensures that both actors interact seamlessly through shared layers, with clear roles and responsibilities. This interplay fosters a feedback loop where human validation improves Generative AI outputs, and Generative AI adapts to human requirements, creating a continuous cycle of enhancement.

A central aspect of the proposed framework is that the ontology is not used solely for post hoc documentation or provenance trace storage, but also to guide Generative AI interaction during task execution. More specifically, tasks, resources, roles, constraints, and contextual attributes are explicitly modeled in the ontology and can be retrieved through SPARQL queries to build structured prompt context for Generative AI systems. As a result, prompts are no longer based only on ad hoc human formulation, but are enriched with task-relevant semantic information derived from the shared collaboration model. This mechanism supports more context-aware and traceable Human--Generative AI collaboration by reducing prompt under-specification and by linking generated outputs to the collaboration elements that informed them.


In summary, the proposed Generative AI Collaboration Framework provides a structured and adaptive approach to leveraging human and Generative AI capabilities. By integrating functional layers and dynamic interactions, it ensures context-aware, efficient, and ethical collaboration across diverse domains. To enhance its ability to manage evolving contexts and interconnected resources, a contextual collaboration AI ontology is essential. In the next section, we explore the development of this ontology and its role in advancing Human-Generative AI collaboration.

\subsection{Development of Contextual Collaboration AI Ontology}
The \textbf{C}ontextual \textbf{C}ollaboration Ontology (CCAI) bridging Human and Generative \textbf{AI}  has been developed to address the need for structured, adaptive, and semantically rich representations of collaboration processes involving human and Generative AI agents. Its design builds upon foundational principles from the PROV-O and FOAF ontologies \cite{lebo2013prov,brickley2014foaf}, complemented by domain-specific extensions to capture dynamic roles, activities, contexts, and resources. Therefore, in this section, we present the development of the ontology, including the definition of objectives, scenarios, competency questions, iterative modeling, and refinement phases to ensure alignment with real-world applications and theoretical foundations.

\subsubsection{Preliminaries, Goals, and Scope of Ontology Development}
The primary goal of the CCAI ontology is to provide a formal framework for modeling collaboration processes involving Human and Generative AI agents based on the principles of SAMOD methodology~\cite{peroni2017simplified}. Its core objectives include representing the dynamics of collaboration by defining the roles, activities, and interactions between agents, whether they are human collaborators or Generative AI systems acting as tools, partners, or autonomous agents~\cite{li2022collaboration, le2023corec}. The CCAI ontology is designed to enhance context-awareness by capturing temporal, spatial, and domain-specific contextual dimensions, ensuring that collaborative processes are well-suited to their unique environments. Additionally, it emphasizes traceability by incorporating provenance mechanisms to track contributions, decisions, and outputs, fostering transparency and accountability. Adaptability across diverse application domains is another critical focus, making the ontology applicable to contexts such as creative industries, decision-making, and problem-solving. 

\begin{table}
	\centering
		\caption{Examples of Key Terms in the Glossary of the CCAI Ontology.}
	\label{table_glossary}
	\begin{tabular}{p{1.2cm} p{7cm} p{3.5cm}}
		\toprule
		\textbf{Term} & \textbf{Definition} & \textbf{Examples} \\\midrule
		\textbf{Agent} & An entity participating in collaboration, including human collaborators, Generative AI systems, and agent groups such as dynamic teams or static units. & Human project manager, AI assistant, research team. \\\midrule
		\textbf{Task} & A specific unit of work within a collaboration process, involving objectives, inputs, and outputs. & Drafting a report, data analysis, product design. \\\midrule
		\textbf{Context} & The domain, spatial, or temporal setting in which collaboration occurs, providing parameters for decision-making. & Education domain, online meeting room, project timeline. \\\midrule
		\textbf{Resource} & Tools, datasets, knowledge bases, or other assets that support collaboration processes. & Text corpora, AI tools, ontologies. \\ \bottomrule
	\end{tabular}
\end{table}

To ground the CCAI ontology in practical applications, scenarios were developed to illustrate interactions between human and Generative AI agents across diverse domains, highlighting the ontology's capacity to model dynamic collaboration processes, roles, and resources. In project management, human project managers collaborate with Generative AI agents to enhance efficiency and decision-making by defining objectives and milestones while delegating routine tasks to the AI. The AI generates detailed schedules, predicts risks using historical data, and suggests optimal resource allocations, which human team members refine to ensure alignment with organizational goals, improving productivity and reducing manual effort. In education, the ontology supports personalized learning by integrating human instructors and Generative AI systems. Instructors design curricula and evaluate progress, while the AI provides personalized study materials, quizzes, and automated feedback tailored to individual learning styles. This partnership fosters an adaptive educational environment where students achieve better outcomes through tailored support and collaboration.

The glossary is a key part of the CCAI ontology, providing clear definitions to ensure consistency in collaborative processes. As shown in Table \ref{table_glossary}, an Agent refers to any entity involved in collaboration, including human collaborators, Generative AI systems, or groups like teams. These agents perform Tasks, which are specific actions with clear goals, inputs, and outputs, such as writing a report or analyzing data. Collaboration happens within a Context, which sets important boundaries, such as the domain of focus or the environment where the work takes place.

\begin{table}
	\centering
	\caption{Informal Competency Questions for the CCAI Ontology.}
	\label{table:competency_questions}
	\begin{tabular}{p{1.0cm} p{11.0cm}}
		\toprule
		\textbf{ID} & \textbf{Competency Question} \\ \midrule
		CQ1 & Which agents contributed to a specific collaborative artifact? \\ \midrule
		CQ2 & What roles are assigned to a Generative AI agent in a task? \\ \midrule
		CQ3 & What resources were used for a specific task? \\ \midrule
		CQ4 & Which tasks are linked to a particular collaboration process? \\\midrule 
		CQ5 & What temporal and spatial contextual attributes define a given collaboration context? \\\midrule
		CQ6 & What ethical constraints apply to this collaboration? \\ \bottomrule
	\end{tabular}
\end{table}

Competency Questions (\textit{CQ}) are essential for guiding the design and scope of the CCAI Ontology, ensuring it addresses relevant use cases effectively. These questions evaluate the ontology's capability to represent agents, tasks, contexts, and their interrelationships within collaborative environments. As listed in Table \ref{table:competency_questions}, agent-related questions examine the roles and contributions of Human and Generative AI agents, such as determining which agents contributed to a specific artifact or identifying the roles assigned to a Generative AI agent for a particular task. Task-related questions focus on the resources used in specific tasks and their connections to broader collaboration processes. Context-related questions address temporal, spatial, and ethical dimensions, including defining contextual attributes and identifying constraints that influence the collaboration.

\subsubsection{Ontology Development}


We focus on the development of the CCAI ontology, which includes the modelet\footnote{A modelet is a small, modular component of an ontology that represents a specific subset of domain knowledge. It can be independently developed, tested, and refined, enabling a divide-and-conquer approach for incremental ontology construction.} development process. This process involves creating preliminary, lightweight modules that serve as prototypes. These modelets help visualize, test, and refine specific aspects of the ontology, such as collaboration contexts, agent roles, or resource relationships, prior to full-scale development. By emphasizing key concepts and their relationships early, modelets ensure that the ontology's structure aligns with the core competency questions and stakeholder requirements.

The primary purpose of modelet development in the CCAI ontology is to construct a skeleton ontology that addresses fundamental collaboration scenarios while remaining adaptable for further refinement. Each modelet defines basic classes, properties, and relationships specific to a particular aspect of the ontology. For example, a modelet may focus on modeling agent roles and competences or on defining the interaction between collaboration contexts and resources. These modules are then tested and validated to ensure they meet competency questions and use case requirements. Once validated, the modelets are integrated into the overarching ontology to form a cohesive and semantically rich representation of contextual collaboration.

Although the goal of this work is not to propose a new ontology engineering methodology, the development of CCAI follows principles that are consistent with established ontology engineering approaches. In particular, our use of modelets is aligned with modular and pattern-oriented development strategies such as eXtreme Design, where smaller reusable modeling units are developed, tested against competency questions, and progressively integrated into a broader ontology network \cite{blomqvist2016engineering}. Similarly, the iterative refinement of requirements, scenarios, glossary terms, competency questions, modeling, and validation resonates with the scenario-based perspective of NeOn \cite{suarez2015neon}, the modular emphasis of Modular Ontology Modeling (MoMo) \cite{shimizu2023modular}, and the specification--conceptualization--validation logic of METHONTOLOGY \cite{lopez1997methontology}. In this sense, CCAI should be understood as an ontology developed through a pragmatic combination of reuse, modularization, iterative validation, and competency-question-driven refinement, rather than as an isolated modeling exercise.

The CCAI ontology is built upon two primary modelets, each focusing on distinct core elements of collaboration. 
The first modelet emphasizes the roles, competences, and activities of Human and Generative AI agents, capturing their individual and collective contributions to collaboration. The second modelet centers on collaboration processes, contexts, and resources, providing a framework to model the settings, tools, and workflows that enable effective collaboration. Together, these modelets form the foundational building blocks of the ontology, facilitating incremental development and integration.

\begin{figure}[h]
	\centering	
	\includegraphics[width=0.98\textwidth]{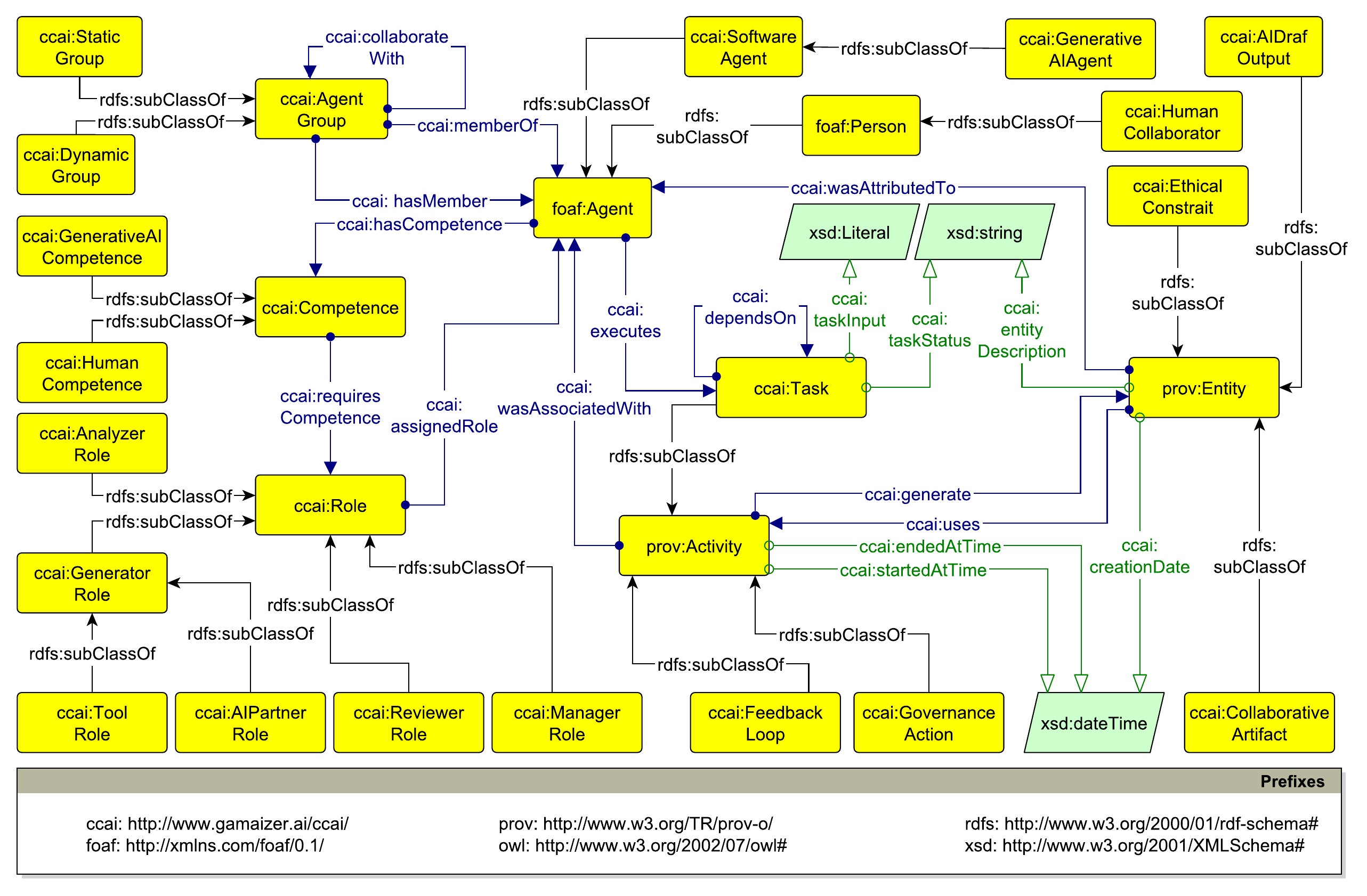}
	
	\caption{Conceptual Modelet of Agents, Activities, Tasks, Competences, and Roles in the CCAI Ontology.}
	\label{fig_03}
\end{figure}

\paragraph{Collaboration Agents, Tasks, Roles, and Entities}
\vspace{0.2cm}
We present the first modelet of the CCAI ontology, which includes the entities, roles, competences, tasks, activities, and entities that form the foundation of collaborative processes involving human and Generative AI agents, as illustrated in Figure~\ref{fig_03}. At its core, the ontology employs the \textit{foaf:Agent} superclass to represent all collaborative participants, encompassing both \textit{ccai:HumanCollaborator} and \textit{ccai:GenerativeAIAgent}. Human collaborators contribute contextual understanding, creativity, and ethical oversight, often assuming leadership, evaluation, or refinement roles. In contrast, Generative AI agents provide specialized competences, such as generating text, synthesizing images, or analyzing data, enhancing productivity by automating routine tasks and generating diverse creative outputs. To account for group-based collaboration, the ontology introduces \textit{ccai:AgentGroup}, which includes \textit{ccai:StaticGroup}, representing fixed membership teams, and \textit{ccai:DynamicGroup}, which evolves based on task requirements or real-time needs. Together, these agent classes enable the ontology to represent both individual and collective contributions to collaborative workflows.

Roles are central to defining the responsibilities and behaviors expected of agents within the collaboration process. General roles include \textit{ccai:AnalyzerRole}, which focuses on data analysis and insights; \textit{ccai:GeneratorRole}, for content or idea creation; \textit{ccai:ManagerRole}, for overseeing planning and execution; and \textit{ccai:ReviewerRole}, for ensuring output quality and alignment with objectives. Specialized roles for Generative AI agents include \textit{ccai:ToolRole}, where AI functions as an assistive tool, and \textit{ccai:AIPartnerRole}, where AI acts as a co-creator, contributing creative and analytical insights. These roles are linked to agents through the \textit{ccai:assignedRole} property, enabling the ontology to effectively model responsibility distribution within collaborative tasks.

Competences capture the specific skills and capabilities that agents bring to collaboration. \textit{ccai:HumanCompetence} represents human expertise in areas like strategic thinking, problem-solving, and creative direction, while \textit{ccai: GenerativeAICompetence} models the advanced capabilities of Generative AI agents, such as natural language processing, machine learning, and image synthesis. The \textit{ccai:hasCompetence} property connects agents to their competences, ensuring alignment between their roles, abilities, and assigned tasks. This alignment supports efficient task execution and optimal resource utilization.

The ontology extends the \textit{prov:Activity} class to represent collaborative processes via \textit{ccai:Task}, which captures individual work units like drafting, reviewing, or editing. Tasks serve as the building blocks of the collaboration process, linking agents, resources, and outputs. Additionally, \textit{ccai:FeedbackLoop} represents iterative exchanges of feedback between agents, facilitating refinement and improvement of outputs, while \textit{ccai:GovernanceAction} models actions taken to enforce ethical guidelines and ensure compliance with governance standards. Tasks are interconnected through properties like \textit{ccai:dependsOn}, which models task dependencies and execution sequences, ensuring that the ontology can accurately represent complex workflows and dynamic interactions.

Specific entities within the ontology represent outputs and constraints that arise during collaboration. \textit{ccai: AIDraftOutput} captures intermediate outputs from Generative AI agents, such as text drafts or image prototypes. \textit{ccai:CollaborativeArtifact} represents final outputs created through collaboration, including documents, designs, or reports, reflecting contributions from both human and AI agents. Finally, \textit{ccai:EthicalConstraint} defines the ethical guidelines and standards governing collaboration, ensuring accountability and compliance. These entities, along with their relationships to agents and activities, provide a framework for provenance tracking, enabling transparency and accountability in collaborative processes.

\begin{figure}[h]
	\centering
	\includegraphics[width=0.98\textwidth]{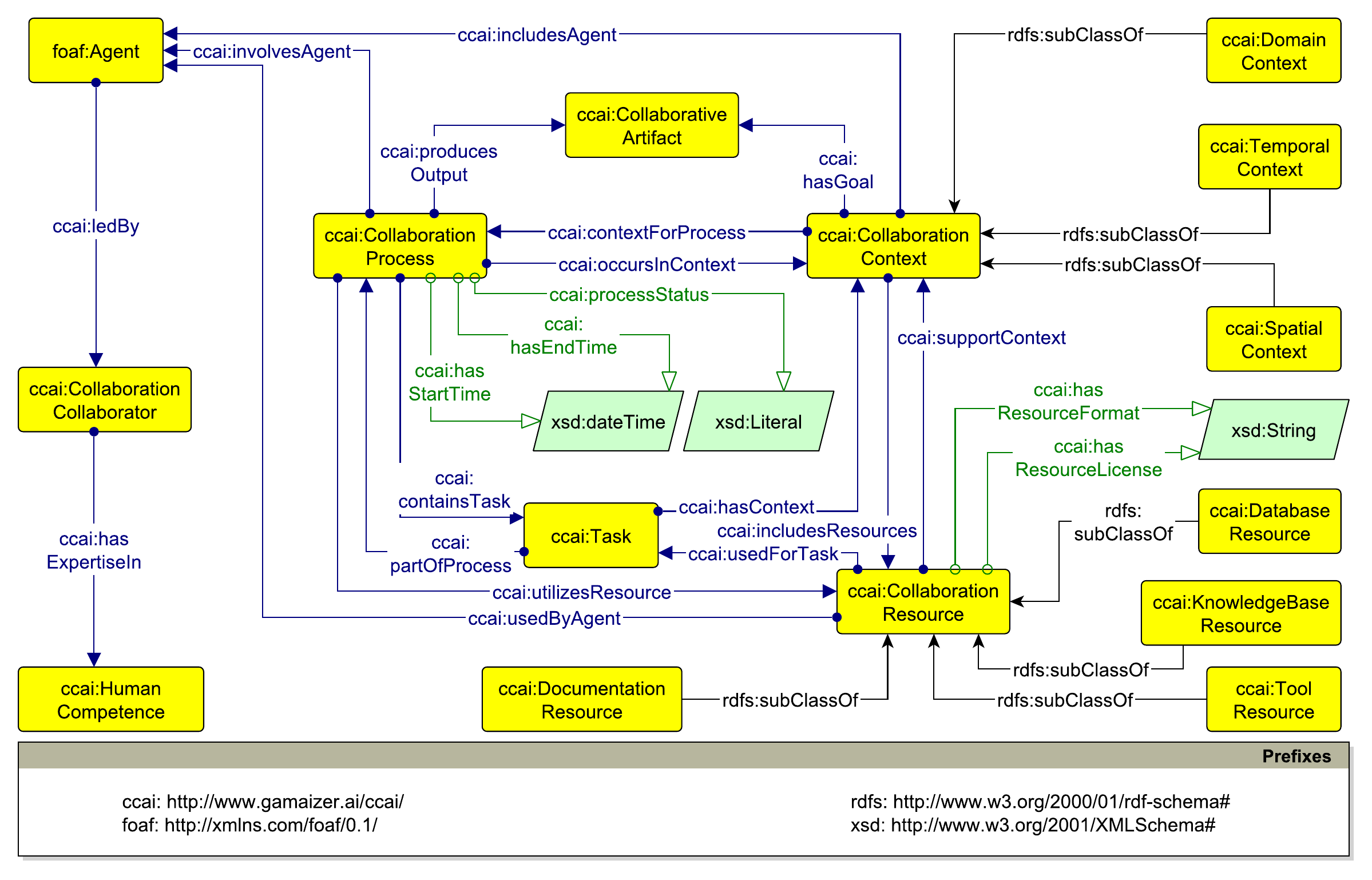}
	\caption{Conceptual modelet of collaboration processes, contexts, resources, tasks, and produced artifacts in the CCAI ontology.}
	\label{fig_04}
\end{figure}

\paragraph{Collaboration Processes, Contexts, Resources, Agents and Tasks}
\vspace{0.2cm}
In the second modelet, we focus on broader aspects of the framework for Human--Generative AI collaboration. This modelet details the relationships between collaboration processes, contexts, resources, and tasks, providing a structured view of how collaborative workflows can be organized and monitored, as shown in Figure~\ref{fig_04}. Accordingly, collaboration processes are represented by the \textit{ccai:CollaborationProcess} class, which captures the overarching workflow of collaborative efforts. A process is composed of multiple tasks through \textit{ccai:containsTask} (with the inverse relation \textit{ccai:partOfProcess}) and integrates participating agents through \textit{ccai:includesAgent} (and \textit{ccai:involvesAgent}). In addition, a process can \textit{ccai:producesOutput} as a \textit{ccai:CollaborativeArtifact}. The execution state of a process is further described by datatype attributes such as \textit{hasStartTime}, \textit{hasEndTime} (\textit{xsd:dateTime}), and \textit{processStatus} (\textit{xsd:Literal}).

The collaboration context defines the setting in which collaboration takes place, encompassing critical parameters such as thematic domains or  timeframes. In Figure~\ref{fig_04}, this is captured through \textit{ccai:CollaborationContext} together with three context types: \textit{ccai:DomainContext}, \textit{ccai:TemporalContext}, and \textit{ccai:SpatialContext}. At the workflow level, a collaboration process is linked to its context using \textit{ccai:occursInContext}, while \textit{ccai:contextForProcess} supports navigation from a context back to its associated process. Moreover, tasks are situated within their context through \textit{ccai:hasContext}, allowing each task to be interpreted with respect to the collaboration setting.

Collaboration resources form another essential component of this modelet, representing the tools and assets used to facilitate tasks and decision-making. The \textit{ccai:CollaborationResource} class captures diverse resources, including \textit{ccai:ToolResource} (e.g., software applications supporting editing or visualization), \textit{ccai:DocumentationResource} (e.g., specifications and reports), \textit{ccai:DatabaseResource} (e.g., structured datasets), and \textit{ccai:KnowledgeBaseResource} (e.g., ontologies and taxonomies). Resource usage is represented at multiple levels: a collaboration process may \textit{ccai:utilizesResource}, tasks may reference supporting resources via \textit{ccai:includesResources}, and each resource may indicate which task it supports through \textit{ccai:usedForTask}. Resources can also be connected to the context they support using \textit{ccai:supportContext}, and to agents through \textit{ccai:usedByAgent}. Finally, resource metadata is captured via datatype properties such as \textit{ccai:hasResourceFormat} and \textit{ccai:hasResourceLicense} (both \textit{xsd:string}).

The relationships between processes, tasks, contexts, and resources are vital for modeling collaborative workflows. Processes structure collaboration by grouping tasks and agents, tasks are anchored to their contextual setting via \textit{ccai:hasContext}, and resources are explicitly associated with processes and tasks through \textit{ccai:utilizesResource}, \textit{ccai:includesResources}, and \textit{ccai:usedForTask}. Together, these links provide a coherent representation of collaborative activities, the contexts in which they occur, the resources they depend on, and the artifacts they produce.

This second modelet enhances the CCAI ontology by addressing the broader relationships and dependencies that underpin effective Human--Generative AI collaboration. It ensures that processes, contexts, resources, tasks, and outputs are represented in a manner aligned with practical requirements, enabling scalable and context-aware applications across diverse domains. With this foundational structure in place, we now turn to test case generation and validation, demonstrating how the ontology can be systematically evaluated to verify its completeness, correctness, and adaptability.

\subsubsection{Test Case Generation and Validation }\label{testcase}

To validate the practical applicability of the CCAI ontology, we designed and executed test cases based on the competency questions. These test cases evaluated the ontology's ability to represent and retrieve information about agents, tasks, resources, and contexts in collaborative environments. SPARQL queries were used to test the ontology's structure and relationships, ensuring alignment with real-world requirements.

One of the key test cases addressed \textit{CQ1}, the final report \textit{ccai:finalReport\_P25} is represented as an instance of \textit{ccai:CollaborativeArtifact} (a subtype of \textit{prov:Entity}).
To retrieve the agents responsible for this artifact in the provenance trace, we use the query in Query Box~\ref{query:agent_contribution}. This query retrieves all agents attributed to the report artifact \textit{ccai:finalReport\_P25} via the \textit{prov:wasAttributedTo} relation. Its successful execution confirms that the ontology represents \textit{ccai:finalReport\_P25} as a collaborative artifact and correctly captures responsibility attribution from artifacts to agents, thereby supporting provenance-based traceability of contributions.

\begin{algorithm}[h]
	\makeatletter
	\renewcommand{\ALG@name}{Query Box}
	\makeatother
	\caption{Retrieve agents contributing to a report.}
	
	\begin{algorithmic}
		\State \textbf{Input} URI of the target artifact (\texttt{ccai:finalReport\_P25})
		\State \textbf{Output} Pairs (\texttt{?artifact}, \texttt{?agent}) where \texttt{?agent} is attributed to \texttt{?artifact}
		
\begin{lstlisting}[style=sparqlstyle,belowskip=0pt]
PREFIX ccai: <http://gamaizer.ai/ccai#>
PREFIX prov: <http://www.w3.org/ns/prov#>
SELECT DISTINCT ?artifact ?agent 
WHERE {
 VALUES ?artifact { ccai:finalReport_P25 }
 ?artifact a ccai:CollaborativeArtifact ;
 prov:wasAttributedTo ?agent .
}
\end{lstlisting}
		
	\end{algorithmic}
	\label{query:agent_contribution}
\end{algorithm}

For \textit{CQ2}, Query Box~\ref{query:genai_roles_in_task} retrieves, for a selected task, each executing agent typed as
\textit{ccai:GenerativeAIAgent} together with the role(s) assigned to that agent. The bindings returned by this query
confirm that the ontology captures explicit role assignment for Generative AI agents during task execution, which
enables role-aware analysis of how AI responsibilities are distributed across collaborative workflows.

\begin{algorithm}[h]
	\makeatletter
	\renewcommand{\ALG@name}{Query Box}
	\makeatother
	\caption{CQ2: Retrieve roles of Generative AI agents executing a task.}
	\begin{algorithmic}
		\State \textbf{Input} Task URI (\texttt{ccai:InitiationAndContextSetting})
		\State \textbf{Output} Tuples (\texttt{?task}, \texttt{?agent}, \texttt{?role})
		\begin{lstlisting}[style=sparqlstyle,belowskip=0pt]
PREFIX ccai: <http://gamaizer.ai/ccai#>
SELECT DISTINCT ?task ?agent ?role 
WHERE {
 VALUES ?task { ccai:InitiationAndContextSetting }
 ?agent a ccai:GenerativeAIAgent ;
 ccai:executes ?task ;
 ccai:assignedRole ?role .
}\end{lstlisting}
	\end{algorithmic}
	\label{query:genai_roles_in_task}
\end{algorithm}

A task can be associated with the collaboration resources it relies on through the \textit{ccai:usedForTask} relation, which links each \textit{ccai:CollaborationResource} to the task it supports. Query Box~\ref{query:task_resources} operationalizes \textit{CQ3} by retrieving, for the selected task, all collaboration resources that declare \textit{ccai:usedForTask} to that task. Successful execution of this query confirms that the ontology captures resource usage and supports analysis of resource dependencies in collaborative workflows.

\begin{algorithm}[h]
	\makeatletter
	\renewcommand{\ALG@name}{Query Box}
	\makeatother
	\caption{CQ3: Retrieve resources used for a task.}
	\begin{algorithmic}
		\State \textbf{Input} Task URI (\texttt{ccai:InitiationAndContextSetting})
		\State \textbf{Output} Pairs (\texttt{?task}, \texttt{?resource})
		\begin{lstlisting}[style=sparqlstyle,belowskip=0pt]
PREFIX ccai: <http://gamaizer.ai/ccai#>			
SELECT DISTINCT ?task ?resource 
WHERE {
 VALUES ?task { ccai:InitiationAndContextSetting }
  ?resource ccai:usedForTask ?task .
  OPTIONAL { ?resource a ccai:CollaborationResource . }
}\end{lstlisting}
	\end{algorithmic}
	\label{query:task_resources}
\end{algorithm}

To operationalize \textit{CQ4}, we retrieve the tasks that belong to a given collaboration process instance. A process may enumerate its constituent tasks through \textit{ccai:containsTask}, while tasks may also explicitly reference their parent process through \textit{ccai:partOfProcess}. Query Box~\ref{query:process_tasks} returns all tasks associated with the selected process (\textit{ccai:ProjectInitiationProcess}). The returned bindings confirm that collaborative activities can be organized and inspected at the process level.

\begin{algorithm}[h]
	\makeatletter
	\renewcommand{\ALG@name}{Query Box}
	\makeatother
	\caption{CQ4: Retrieve tasks linked to a collaboration process.}
	\begin{algorithmic}
		\State \textbf{Input} Collaboration process URI (\texttt{ccai:ProjectInitiationProcess})
		\State \textbf{Output} Tuples (\texttt{?process}, \texttt{?task}, \texttt{?taskName})
		\begin{lstlisting}[style=sparqlstyle,belowskip=0pt]
PREFIX ccai: <http://gamaizer.ai/ccai#>			
SELECT DISTINCT ?process ?task ?taskName 
WHERE {
 VALUES ?process { ccai:ProjectInitiationProcess }
 { ?process a ccai:CollaborationProcess ;
  ccai:containsTask ?task . }
 UNION
  { ?task a ccai:Task ;
   ccai:partOfProcess ?process . }			
 OPTIONAL { ?task ccai:taskName ?taskName . }
}\end{lstlisting}
	\end{algorithmic}
	\label{query:process_tasks}
\end{algorithm}

To operationalize \textit{CQ5}, we retrieve the temporal and spatial descriptors that characterize collaboration contexts. Query Box~\ref{query:context_temporal_spatial} enumerates all instances of \textit{ccai:CollaborationContext} and, when available, extracts their domain label as well as temporal (\textit{ccai:startDate}, \textit{ccai:endDate}) and spatial (\textit{ccai:locationName}) attributes. These attributes are stored directly on the context instance, which may also be typed as \textit{ccai:DomainContext}, \textit{ccai:TemporalContext}, and/or \textit{ccai:SpatialContext}. The returned bindings confirm that the ontology supports an explicit, queryable representation of collaboration context along domain, time, and space dimensions.

\begin{algorithm}[h]
	\makeatletter
	\renewcommand{\ALG@name}{Query Box}
	\makeatother
	\caption{CQ5: Retrieve temporal \& spatial attributes of collaboration contexts.}
	\begin{algorithmic}
		\State \textbf{Input} None (enumerates all \texttt{ccai:CollaborationContext} instances)
		\State \textbf{Output} Tuples (\texttt{?context}, \texttt{?domainLabel}, \texttt{?start}, \texttt{?end}, \texttt{?location})
		
		\begin{lstlisting}[style=sparqlstyle,belowskip=0pt]
PREFIX ccai: <http://gamaizer.ai/ccai#>
SELECT DISTINCT ?context ?domainLabel ?start ?end ?location 
WHERE {
 ?context a ccai:CollaborationContext .	
 OPTIONAL {
  ?context a ccai:DomainContext .
  OPTIONAL { ?context ccai:domainLabel ?domainLabel . }
 }	
 OPTIONAL {
  ?context a ccai:TemporalContext .
  OPTIONAL { ?context ccai:hasStartDate ?start . }
  OPTIONAL { ?context ccai:hasEndDate ?end . }
 }	
 OPTIONAL {
  ?context a ccai:SpatialContext .
  OPTIONAL { ?context ccai:locationName ?location . }
 }
}\end{lstlisting}
	\end{algorithmic}
	\label{query:context_temporal_spatial}
\end{algorithm}

To answer \textit{CQ6}, we retrieve the ethical constraints associated with collaboration contexts in the knowledge base. Query Box~\ref{query:ethical_constraints} enumerates all instances of \textit{ccai:CollaborationContext} and, when available, returns the constraint entities linked to each context. This supports compliance inspection across collaborations by making constraints explicitly queryable.

\begin{algorithm}[h]
	\makeatletter
	\renewcommand{\ALG@name}{Query Box}
	\makeatother
	\caption{CQ6: Retrieve ethical constraints applicable to collaboration contexts.}
	\begin{algorithmic}
		\State \textbf{Input} None (enumerates all \texttt{ccai:CollaborationContext} instances)
		\State \textbf{Output} Tuples (\texttt{?context}, \texttt{?constraint}, \texttt{?constraintLabel})
		\begin{lstlisting}[style=sparqlstyle,belowskip=0pt]
PREFIX ccai: <http://gamaizer.ai/ccai#>			
SELECT DISTINCT ?context ?constraint ?constraintLabel
WHERE {
 ?context a ccai:CollaborationContext .
 OPTIONAL {
  ?context ccai:hasEthicalConstraint ?constraint .
  OPTIONAL { 
   ?constraint ccai:constraintLabel ?constraintLabel . 
  }
 }
}\end{lstlisting}
	\end{algorithmic}
	\label{query:ethical_constraints}
\end{algorithm}

Through these queries, we validated the CCAI ontology's capability to represent collaboration processes and retrieve relevant information. These tests ensured that the ontology aligns with its intended use cases and competency questions. Iterative refinements based on the outcomes further enhanced its semantic precision and adaptability, making it robust for diverse collaborative scenarios. Building on this validated foundation, we now turn to a case study demonstrating how the ontology applies to a collaborative project management setting.

\subsubsection{Ontology Evaluation}
\label{sec:ontology-evaluation}

We evaluated CCAI along three complementary dimensions: (i) \emph{structural completeness}, (ii) \emph{logical consistency}, and (iii) \emph{competency-question} (CQ) coverage. Structural metrics were obtained with  \textit{OntoMetrics} ~\cite{lozano2004ontometric}, consistency was verified using a Description Logic reasoner (HermiT) \cite{glimm2014hermit}, and CQ coverage was assessed via SPARQL over the illustrative ABox. The ontology is publicly available.\footnote{\url{https://github.com/lengocluyen/ccai_ontology}}

\paragraph{Structural metrics: }
Table~\ref{tab:ontometrics} summarises OntoMetrics results for CCAI. Overall, CCAI exhibits a moderate size and balanced use of classes and properties, with DL expressivity $\mathcal{ALH(D)}$ (attributive language with role hierarchies and datatypes).

\begin{table}[ht]
	\centering
	\caption{OntoMetrics summary for the CCAI Ontology.}
	\label{tab:ontometrics}
	\begin{tabular}{l r}
		\toprule
		\textbf{Base metrics} & \\
		\midrule
		Axioms (total) & 196 \\
		Logical axioms & 97 \\
		Classes / Total classes & 22 / 49 \\
		Object properties / Total & 11 / 52 \\
		Data properties / Total & 3 / 9 \\
		Individuals / Total & 25 / 29 \\
		DL expressivity & $\mathcal{ALH(D)}$ \\
		\midrule
		\textbf{Class axioms} & \\
		\midrule
		\quad SubClassOf & 43 \\
		\quad DisjointClasses & 4 \\
		\midrule
		\textbf{Object property axioms} & \\
		\midrule
		\quad SubObjectPropertyOf & 36 \\
		\quad InverseObjectProperties & 3 \\
		\quad Domain / Range axioms & 53 / 49 \\
		\quad Property chains & 13 \\
		\midrule
		\textbf{Data property axioms} & \\
		\midrule
		\quad Domain / Range axioms & 9 / 8 \\
		\midrule
		\textbf{ABox (individual axioms)} & \\
		\midrule
		\quad Class assertions & 25 \\
		\quad Object property assertions & 27 \\
		\quad Data property assertions & 8 \\
		\midrule
		\textbf{Schema-level indicators} & \\
		\midrule
		Attribute richness & 0.184 \\
		Inheritance richness & 0.878 \\
		Relationship richness & 0.566 \\
		Inverse relations ratio & 0.058 \\
		Class richness (KB) & 0.286 \\
		\bottomrule
	\end{tabular}
\end{table}

The hierarchy is well-formed (inheritance richness $=0.878$) with moderate relational modelling (relationship richness $=0.566$). Attribute richness ($0.184$) and the near-zero attribute-to-class tendency indicate that descriptive information is primarily conveyed via object properties and instances rather than many datatype attributes - appropriate for a provenance/trace ontology emphasising \textit{Task}-\textit{Role}-\textit{Resource} relations. The $\mathcal{ALH(D)}$ profile supports tractable reasoning while accommodating role hierarchies. The number of domain/range axioms (53/49) and 13 property chains reflects deliberate constraint modelling for traceability.

\paragraph{Logical consistency (reasoner): }
HermiT detected no unsatisfiable classes. Subsumptions against PROV-O superclasses (e.g., \textit{ccai:Task}~$\sqsubseteq$~\textit{prov:Activity}) are satisfiable. No conflicts were observed between declared domains/ranges and the example ABox.

\paragraph{Competency Question coverage: }
Six CQs were executed as SPARQL queries and returned the expected bindings over the ABox (see Test Case~\ref{testcase}),
covering: (\textit{CQ1}) agents attributed to a specific collaborative artifact (e.g., \textit{ccai:finalReport\_P25}), (\textit{CQ2}) roles assigned to Generative AI agents executing a given task, (\textit{CQ3}) collaboration resources used for a specific task, (\textit{CQ4}) tasks associated with a particular collaboration process, (CQ5) temporal and spatial contextual attributes that characterize collaboration contexts, and (\textit{CQ6}) ethical constraints applicable to collaboration contexts. This confirms the operational queryability of the intended modelling scope.

\subsection{Framework Evaluation}\label{sec:evaluation}

To demonstrate the feasibility and practical utility of the CCAI-based framework, we present an illustrative case study based on a real software project. The goal of this illustration is not to make generalizable empirical claims, but to demonstrate how the framework's components function in a realistic workflow and to highlight its potential benefits for transparency and accountability.

Our inquiry focused on three questions: (RQ1) how an explicit, shared ontology shapes the workflow between human developers and AI agents; (RQ2) in what ways the framework supports transparency and accountability throughout the development lifecycle; and (RQ3) what practical benefits and challenges arise when implementing an ontology-driven approach.

Related to data sources and procedure, we instantiated the framework in the project, observed collaboration practices, and analyzed resulting artifacts (e.g., code commits, prompts, generated tests, documentation, and the populated ontology). We also issued targeted SPARQL queries to capture context (tasks, roles, resources) and trace provenance, using these traces as evidence when interpreting observations. Having defined our evaluation methodology and research questions, we next present a comprehensive case study to assess the framework's impact on transparency and collaboration.

To strengthen the empirical component while preserving the illustrative character of the study, Table~\ref{tab:empirical_comparison} reports a bounded comparison at two levels: first, a prompt-level comparison for the representative task ``\textit{View \& Update Competency Profiles}'', which is detailed in the case study in Section~4; and second, an aggregate output-level comparison across the 12 task instances currently represented in the populated ABox.

\begin{table}[t]
	\centering
	\caption{Preliminary empirical comparison between prompt-only (PO) and CCAI-backed (CB) conditions. The illustrative-task columns report the prompt-level comparison for ``\textit{View \& Update Competency Profiles}'', while the all-task columns report the aggregate output-level comparison across the 12 task instances currently instantiated in the populated ABox.}
	\label{tab:empirical_comparison}
	\begin{tabular}{@{}p{0.42\textwidth}cccc@{}}
		\toprule
		\multirow{2}{*}{\textbf{Indicator}} &
		\multicolumn{2}{c}{\textbf{Illustrative task}} &
		\multicolumn{2}{c}{\textbf{All tasks (aggregate)}} \\
		\cmidrule(lr){2-3} \cmidrule(lr){4-5}
		& \textbf{PO} & \textbf{CB} & \textbf{PO} & \textbf{CB} \\
		\midrule
		Contextual categories explicit & 0/4 & 4/4 & 1/46 & 46/46 \\
		Resources explicitly named & 0/3 & 3/3 & 1/31 & 31/31 \\
		Role--agent assignments explicitly named & 0/3 & 3/3 & 8/32 & 32/32 \\
		Omitted task-linked contextual items & 8 & 0 & 86/95 & 0/95 \\
		Structured provenance path available & No & Yes & 0/12 & 12/12 \\
		\bottomrule
	\end{tabular}
\end{table}

The columns under ``Illustrative task'' evaluate how much task-relevant context is explicitly present in the constructed prompt for the representative task, whereas the columns under ``All tasks (aggregate)'' evaluate how much of that context is preserved in generated outputs across all 12 instantiated tasks. These two blocks are therefore complementary rather than identical evaluation units.

The indicators are aligned with the central claims of the paper. ``Contextual categories explicit'' evaluates whether the available contextual categories associated with a task, namely context, resources, responsibilities, and constraints, are explicitly represented. ``Resources explicitly named'' and ``Role--agent assignments explicitly named'' assess whether concrete task-linked resources and responsibility assignments retrieved from the ontology are preserved. ``Omitted task-linked contextual items'' captures how many relevant contextual elements are missing, while ``Structured provenance path available'' indicates whether the resulting artifact can be explicitly linked back to the originating task and responsible contributors through provenance relations.

Overall, the results indicate a consistent advantage for the CCAI-backed condition. For the illustrative task, ontology-backed prompting makes all contextual categories, resources, and role--agent assignments explicit, whereas the prompt-only condition leaves them implicit. At aggregate level, the same pattern appears across the 12 instantiated tasks: the CCAI-backed condition preserves contextual structure and provenance information much more consistently, while the prompt-only condition frequently omits task-linked elements. These results provide bounded empirical support for the framework's claims regarding contextual grounding, traceability, and accountability, while remaining narrower than a controlled large-scale evaluation.

\section{Case Study of the CCAI Ontology}
This section presents an illustrative case study demonstrating the application of the Human-Generative AI Collaboration Framework and the CCAI Ontology in a software project management scenario. The goal is not empirical testing but to demonstrate how the ontology captures, represents, and queries provenance and contextual relations within a realistic workflow. The case focuses on developing a competency-based education management system, a representative domain where human expertise and Generative AI agents collaborate across project planning, design, implementation, and monitoring phases. 

\subsection{Project Background and Participants}
The competency-based education management system aims to provide an integrated platform that evaluates employee skills, identifies skill gaps, and recommends personalized training paths~\cite{luyen2025automatedskilldecompositionmeets,le2025llmspredictprerequisiteskills}. The system is developed by a mid-sized software company, where distributed teams  -- including project managers, technical leads, developers, Quality Assurance (QA) engineers, and UX/UI designers  -- work together with Generative AI agents to accelerate development and improve system quality. To ensure transparency and accountability, the entire collaboration process is semantically annotated using the CCAI ontology.

A recurring challenge in prior projects motivated this ontology-first approach. Teams frequently relied on unstructured prompts to drive Generative AI assistance, which led to context drift: syntactically valid yet contextually incorrect outputs. In one representative incident, an AI-generated component for a user profile page assumed a generic schema and ignored the project's actual database fields, requiring a half day of rework by the developer and technical lead. The CCAI-based framework was introduced to help reduce such ambiguity by making task, role, resource, and constraint information explicit and queryable in day-to-day work.

\begin{table}
	\centering
	\caption{Agent Roles in the Competency-based Education Management System.}
	\label{tab:agent_roles}
	\begin{tabular}{@{}p{0.25\textwidth}p{0.18\textwidth}p{0.50\textwidth}@{}}
		\toprule
		\textbf{Collaborator} &\textbf{Role} & \textbf{Description} \\ \midrule
		\multirow{3}{*}{Human Agents} & Project Manager (PM) & Oversees project planning, risk assessment, and resource allocation. \\\cmidrule{2-3}
		&Technical Lead & Guides system architecture and reviews Generative AI-generated outputs. \\\cmidrule{2-3}
		&Developers \& QA Engineers & Implement features, conduct testing, and manage iterative feedback. \\\cmidrule{2-3}
		&UX/UI Designers & Develop the user interface with a focus on user-friendly competency tracking and analytics. \\ \midrule
		
		\multirow{3}{*}{Generative AI Agents} & Generative AI Code Assistant & Functions as a tool to provide code autocompletion, suggestions, and debugging support. \\\cmidrule{2-3}
		& Generative AI Test Generator & Acts as a collaborative partner by generating and refining test cases. \\\cmidrule{2-3}
		& Generative AI Analytics Agent & Operates autonomously to monitor project metrics, trigger alerts, and recommend adjustments throughout development. \\
		\bottomrule
	\end{tabular}
\end{table}
The framework defines distinct roles for both Human and Generative AI agents, which are represented in the ontology as shown in Table~\ref{tab:agent_roles}. In our approach, each role is modeled as a specific ontology class that encapsulates the functional responsibilities and competencies required for effective collaboration. These classes serve as the foundational building blocks for role definition, ensuring that each agent's contributions are clearly categorized. Moreover, the roles are systematically linked to specific tasks through properties such as \textit{ccai:assignedRole} and \textit{ccai:hasCompetence}, which provide a robust semantic mapping between the agents' functions and their operational activities.

\begin{table}[t]
	\centering
	\caption{Concrete Generative AI services used to instantiate the AI roles in the case study.}
	\label{tab:ai_services}
	\begin{tabular}{p{0.22\textwidth}p{0.22\textwidth}p{0.24\textwidth}p{0.18\textwidth}}
		\toprule
		\textbf{AI role} & \textbf{Service / model} & \textbf{Main use} & \textbf{Interaction mode} \\
		\midrule
		Code Assistant & GitHub Copilot & Code completion, suggestions, debugging support & Interactive \\
		Test Generator & Claude (Anthropic) via API & Test-case generation and refinement & Interactive / batch \\
		Analytics Agent & Claude (Anthropic) via API & Metric monitoring, summarization, recommendations & Periodic / event-driven \\
		\bottomrule
	\end{tabular}
\end{table}

To improve replicability, Table~\ref{tab:ai_services} summarizes the concrete Generative AI services used to instantiate these roles in the case study. These services were selected according to their primary usage: GitHub Copilot was used as an interactive coding assistant during implementation, while Claude Sonnet 3.7 was used for test generation and analytics-oriented reasoning tasks through prompt-based interactions. All generated outputs were reviewed by human collaborators before integration into project artifacts.

\subsection{Collaboration Workflow Process}
\begin{figure}[h]
	\centering	
	\includegraphics[width=5.1in]{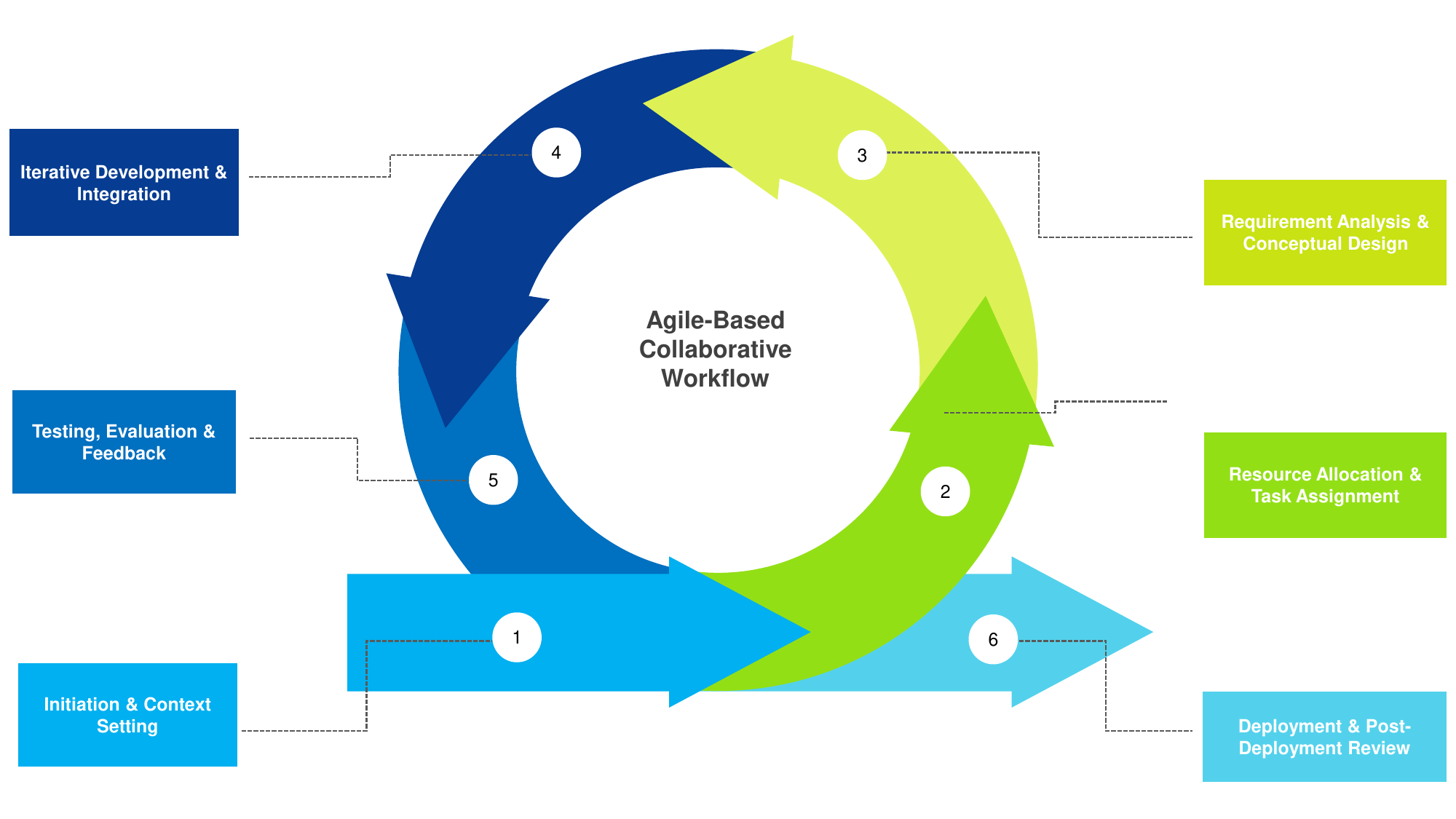}
	\caption{Overall Agile-Based Collaboration Workflow Process.}
	
	\label{fig_05}
\end{figure}

The collaboration workflow process integrates the Human-Generative AI Collaboration Framework with the CCAI ontology to manage projects in a dynamic, transparent, and accountable manner. This workflow encompasses six main phases, as shown in Figure~\ref{fig_05}, each supported by semantic annotations that enable continuous tracking of context, roles, and outcomes throughout the project lifecycle, as summarized in Table~\ref{tab:agile_workflow}.

\begin{figure}[h]
	\centering
	\includegraphics[width=5.1in]{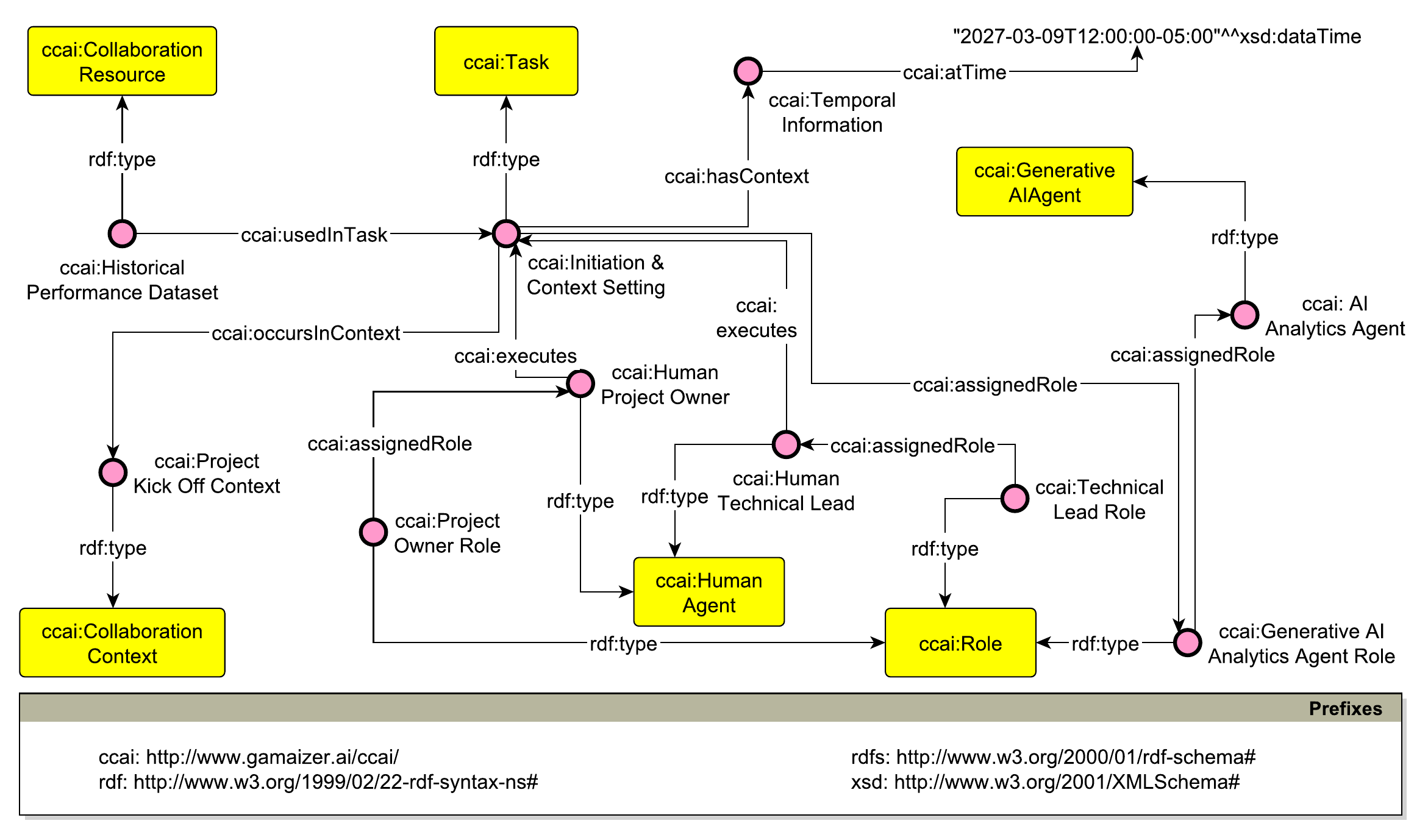}
	\caption{Illustration of instances for the \textit{Initiation \& Context Setting} phase. The main task
		(\textit{ccai:InitiationAndContextSetting}) is typed as \textit{ccai:Task} and linked to a collaboration context
		(\textit{ccai:ProjectKickOffContext}) via \textit{ccai:occursInContext}. The task is additionally connected to temporal
		information (\textit{ccai:TemporalInformation}) through \textit{ccai:hasContext}, with a start date captured by
		\textit{ccai:startedAtTime}. A collaboration resource (\textit{ccai:HistoricalPerformanceDataset}) is typed as
		\textit{ccai:CollaborationResource} and linked to the task through \textit{ccai:usedForTask}. Agents
		(\textit{ccai:HumanProjectOwner}, \textit{ccai:HumanTechnicalLead}, \textit{ccai:AIAnalyticsAgent}) are connected to the
		task via \textit{ccai:executes} and to their roles via \textit{ccai:assignedRole}.}
	\label{fig_08}
\end{figure}

\textbf{\textit{Initiation \& Context Setting}}: 
The work typically starts with a project kickoff that clarifies the overall vision and scope, followed by an initial backlog brainstorming. As illustrated in Figure~\ref{fig_08}, this phase is captured in the CCAI knowledge base by instantiating the main task (\textit{ccai:InitiationAndContextSetting}) as a \textit{ccai:Task} and linking it to its collaboration context (\textit{ccai:ProjectKickOffContext}) via \textit{ccai:occursInContext}. In addition, temporal information is attached to the task through \textit{ccai:hasContext} (e.g., \textit{ccai:TemporalInformation} with \textit{ccai:startedAtTime}), ensuring that time-related constraints are explicitly recorded. During this phase, baseline evidence is incorporated as a collaboration resource (\textit{ccai:HistoricalPerformanceDataset}) and connected to the task via \textit{ccai:usedForTask}. Finally, participants are represented as agents who \textit{ccai:executes} the task and are assigned their respective roles through \textit{ccai:assignedRole}, namely \textit{ccai:ProjectOwnerRole}, \textit{ccai:TechnicalLeadRole}, and \textit{ccai:GenerativeAIAnalyticsAgentRole}. 

\begin{table}
	\centering
	\caption{Agile-Based Collaborative Workflow for Project Development.}
	\label{tab:agile_workflow}
	\begin{tabular}{@{}p{0.15\textwidth} p{0.18\textwidth} p{0.3\textwidth} p{0.27\textwidth}@{}}
		\toprule
		\textbf{Phase} & \textbf{Agile Activities} & \textbf{Key Tasks} & \textbf{Primary Roles} \\ \midrule
		
		Initiation \& Context Setting & 
		Project Kickoff; Initial Product Backlog; CCAI Context Setup
		& 
		Define scope and objectives; Stakeholder analysis; Gather initial requirements; Tag context in ontology
		& 
		PM; Technical Lead; Generative AI Analytics Agent \\\hline
		
		Resource Allocation \& Task Assignment &
	 	Backlog Refinement; Initial Sprint Planning
		& Decompose epics into user stories; Assign roles with \textit{ccai:assignedRole};
			Allocate dev/design resources
		&
			PM; Technical Lead; Developers; Generative AI Code Assistant\\\hline
		
		Requirements Analysis \& Conceptual Design & User Story Elaboration; Backlog Grooming & Define acceptance criteria; Draft system design/UI; Use Generative AI insights for innovative patterns 
		& Business Analyst; Domain Expert; Technical Lead; Generative AI Design Assistant \\\hline
		
		Iterative Development \& Integration & 
			Sprint Execution; Daily Stand-ups; Continuous Integration
		&
			Implement features; Generative AI-assisted code generation; Incorporate feedback
			(\textit{ccai:FeedbackLoop})
		&
			Developers; QA Engineers; Generative AI Code Assistant; Generative AI Test Generator
		\\\hline
		
		Testing, Evaluation \& Feedback & Sprint Review; Sprint Retrospective
		& Execute test suites; Monitor metrics; Update backlog with defects
		& QA Engineers; Generative AI Test Generator; Generative AI Analytics Agent
		\\\hline
		
		Deployment \& Post-Deployment Review &
		Release Planning; Production Deployment
		& System deployment; User training; Document lessons; Provenance tracking
		& Deployment Specialist; PM; Technical Lead; Generative AI Analytics Agent
		 \\ \bottomrule
		
	\end{tabular}
\end{table}
\textbf{\textit{Resource Allocation \& Task Assignment}}:
Once the high-level objectives are agreed upon, the team decomposes epics into more granular user stories, assigning them to both human and Generative AI collaborators. This decomposition is refined through backlog refinement sessions and initial sprint planning, where the team estimates story points, prioritizes backlog items, and plans capacity for the upcoming iterations. The CCAI plays a pivotal role in this process by offering properties such as \textit{ccai:assignedRole} and \textit{ccai:hasCompetence} to ensure that tasks are matched with the appropriate agents. Key tasks involve defining specific coding, testing, or design requirements, allocating resources accordingly, and updating the ontology to reflect task dependencies (e.g., \textit{ccai:dependsOn}). The primary human roles here include the Product Owner/PM, the Technical Lead, and Developers, while the AI Code Assistant provides automated support for assigned tasks.

\textbf{\textit{Requirements Analysis \& Conceptual Design}}:
During this phase, the team refines the user stories by delving into detailed acceptance criteria and discussing the technical and conceptual architecture needed for the project. Activities often include user story elaboration, where acceptance criteria are clarified, and architecture or UI/UX prototyping sessions that produce wireframes and technical diagrams. Backlog grooming runs in parallel, continuously adjusting and reordering items based on newly discovered insights. The key tasks here involve documenting detailed functional and non-functional requirements, leveraging Generative AI Design Assistant capabilities to create interface mockups or architectural suggestions, and semantically tagging all design decisions in the ontology (for instance, using \textit{ccai:AIDraftOutput}). Primary human roles during this stage include the Product Owner or Business Analyst, the Technical Lead, and the UX/UI Designer, working closely with the Generative AI Design Assistant to ensure an innovative yet coherent solution blueprint.

\textbf{\textit{Iterative Development \& Integration}}:
Having established a clear design approach, the team moves into a series of iterative sprints to build the system incrementally. Developers are responsible for creating or refining features, guided by user stories acceptance criteria, while Generative AI agents assist with code generation and test creation. During these sprints, daily stand-ups keep everyone aligned on progress and obstacles, and continuous integration ensures that code is frequently merged and validated against build and test pipelines. Key tasks include the implementation of prioritized features, applying refactoring suggestions from the AI Code Assistant, and generating automated test scripts and coverage analysis through the Generative AI Test Generator. Feedback loops captured in \textit{ccai:FeedbackLoop} allow both human and Generative AI contributors to refine the output incrementally. Developers and QA Engineers take the lead in building and validating the system, supported by the Generative AI Code Assistant and Generative AI Test Generator to streamline coding and testing efforts.

\textbf{\textit{Testing, Evaluation \& Continuous Feedback}}:
Testing is woven throughout the entire sprint process, but it becomes a pronounced focus as user stories near completion. QA Engineers and the Generative AI Test Generator cooperate to execute functional, performance, and regression tests, identifying defects that are then tracked in the backlog for resolution. The team conducts sprint reviews to demonstrate completed user stories to stakeholders, gathering feedback that helps evaluate whether the increment meets expectations. Sprint retrospectives follow, where the process itself is critiqued, and improvements are suggested for subsequent cycles. Additional support from the Generative AI Analytics Agent enables the team to monitor metrics such as velocity and risk indicators, helping them make data-driven decisions. The ontology is continually updated with test outcomes, bug fixes, and iteration results, ensuring that every change is systematically documented.

\textbf{\textit{Deployment \& Post-Deployment Review}}:
Once the final sprint validates the feature set, the system is deployed to the production environment, concluding the development cycle. The Deployment Specialist collaborates with the PM or Scrum Master and the Technical Lead to finalize the release scope, plan the transition strategy, and manage any necessary data migration. Post-deployment, the team gathers user input and operational insights to guide potential improvements or patches. Lessons learned are formally documented, and provenance information is appended to the ontology, for example using \textit{prov:wasAttributedTo} to track agent contributions. At this stage, the Generative AI Analytics Agent continues to monitor performance metrics, revealing how well the system aligns with real-world conditions and offering opportunities for iterative enhancements.

Having outlined the six main phases of the collaboration workflow process in a case study of project management  -- from Initiation \& Context Setting through Deployment \& Post-Deployment Review -- we now delve into a specific task to illustrate Human-Generative AI collaboration in action. In the following section, we present a concrete example of how human agents and Generative AI agents cooperate, leveraging the CCAI ontology to accomplish a particular development objective.

\subsection{Collaborative Task Implementation}
In our framework, prompt construction is semi-automatic. A human collaborator first initiates a request associated with a specific task instance, after which task-relevant contextual information is retrieved from the ontology through SPARQL queries and incorporated into the final prompt sent to the Generative AI system. The resulting prompt therefore combines user intent with ontology-derived contextual information, allowing the generated response to remain aligned with the collaboration process, the available resources, the responsible roles, and the execution context.

\begin{figure}[h]
	\centering
	\includegraphics[width=5.1in]{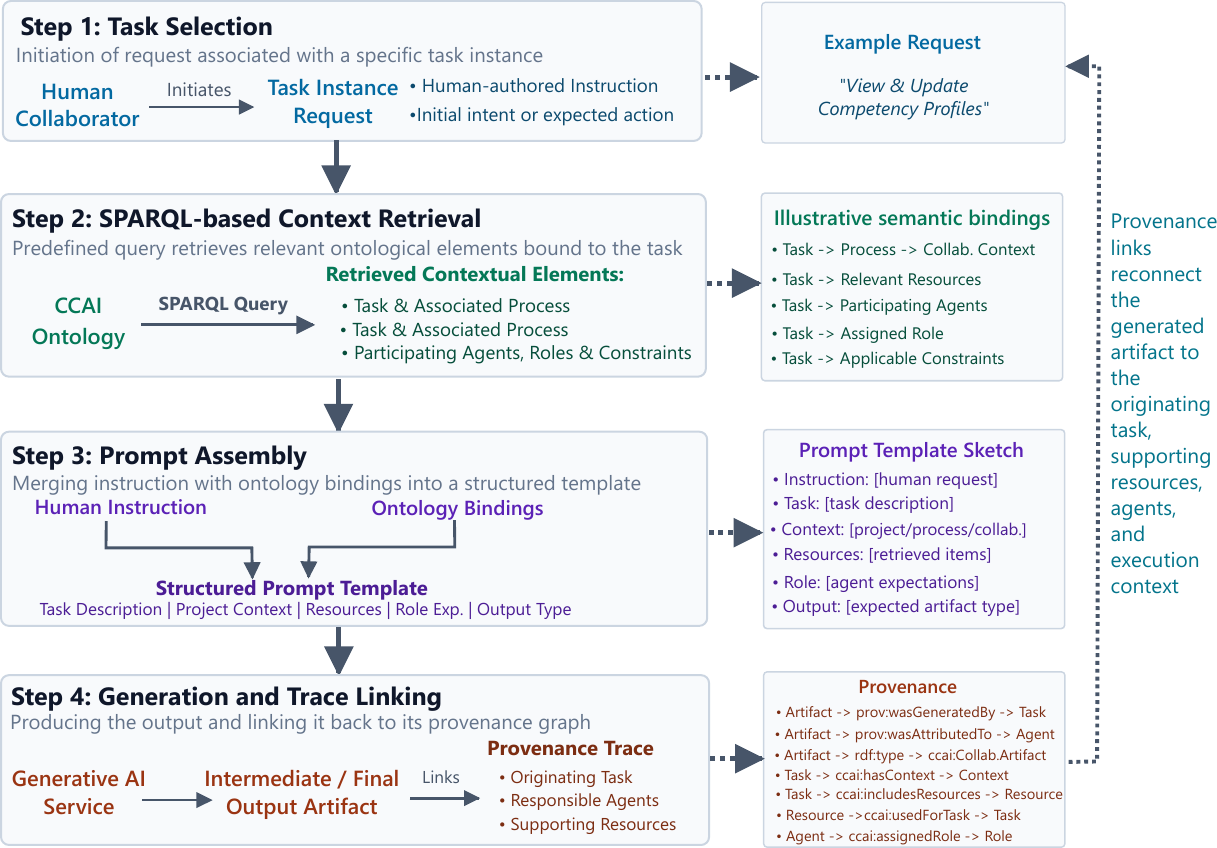}
	\caption{SPARQL-driven pipeline for context-aware prompt construction and provenance-linked artifact generation.}
	\label{fig:sparql_prompt_pipeline}
\end{figure}

To make this process explicit, and as shown in Figure~\ref{fig:sparql_prompt_pipeline}, we model prompt construction as a four-step SPARQL-driven pipeline. Step~1 is task selection: a human collaborator formulates a request associated with a concrete task instance in the project backlog. Step~2 is SPARQL-based context retrieval: a predefined query is executed over the CCAI ontology to retrieve the contextual elements associated with that task, including the collaboration process, execution context, relevant resources, participating agents, and assigned roles. Step~3 is prompt assembly: the human-authored instruction is merged with the retrieved ontology bindings in a structured template whose fields explicitly capture the task description, project context, supporting resources, role expectations, and expected output type. Step~4 is generation and trace linking: the resulting prompt is submitted to the Generative AI service, and the produced output is linked back to the originating task and related collaboration entities through provenance-aware relations.

\begin{figure}[h]
	\centering	
	\includegraphics[width=5.1in]{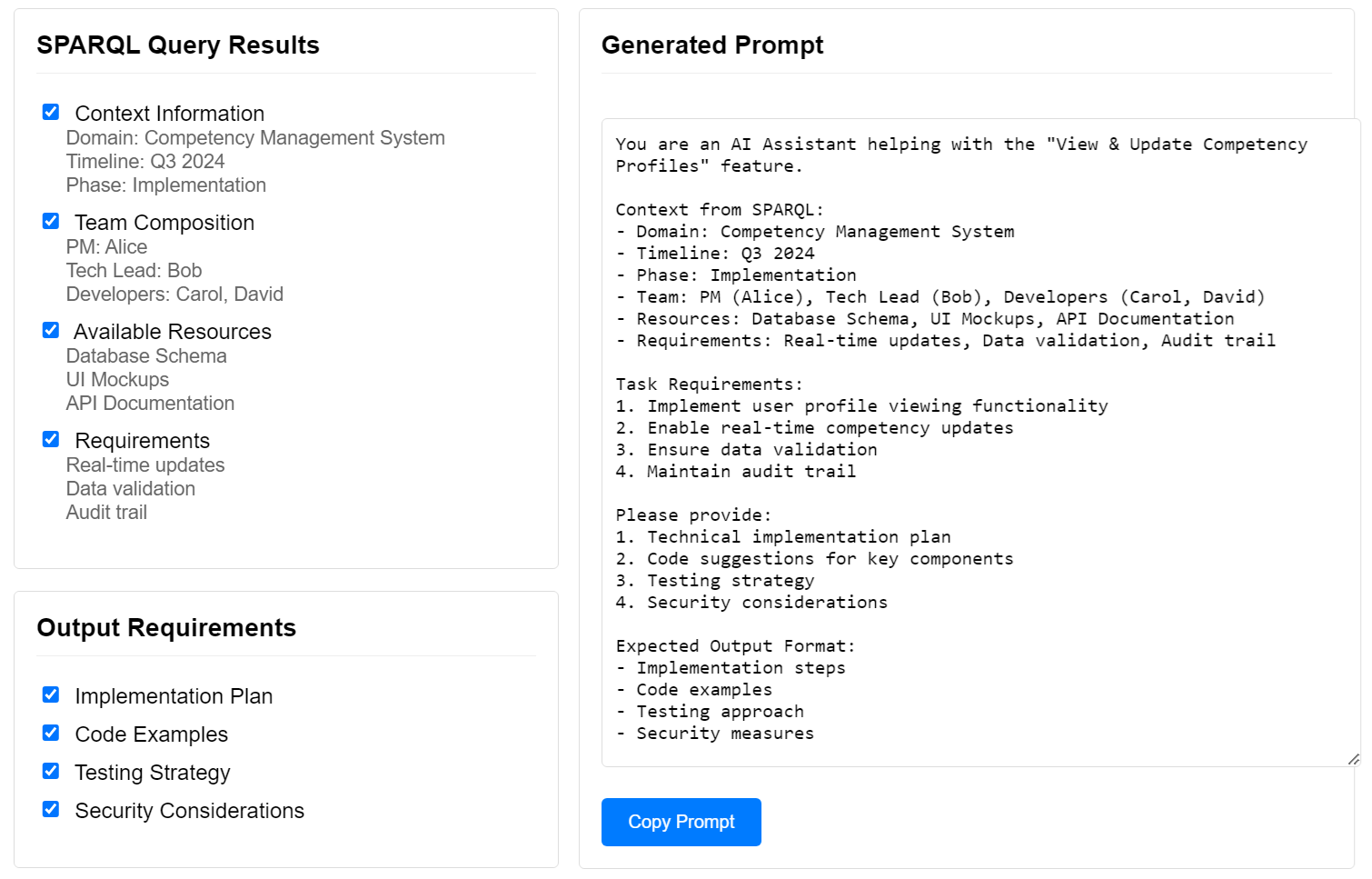}
	
	\caption{Generative AI Prompt template for the ``\textit{View \& Update Competency Profiles}'' Task.}
	\label{fig_07}
\end{figure}

We now focus on a concrete task in the project backlog: enabling registered users to view and update their competency profiles.
This task is implemented during a sprint (\textit{ccai:Sprint1Context}) and requires coordination across development, AI-assisted coding, and QA.
Figure~\ref{fig:sparql_prompt_pipeline} summarizes the end-to-end workflow that is instantiated below through Query Box~\ref{query:query_ex}, the returned bindings in Table~\ref{tab:query_view_update_results} (Step~2), and the prompt template in Figure~\ref{fig_07} (Step~3).
\begin{algorithm}[h!]
	\makeatletter
	\renewcommand{\ALG@name}{Query Box}
	\makeatother
	\caption{Retrieve context, resources, process, roles, and agents for the \texttt{``View \& Update Competency Profiles''} task}
	\begin{algorithmic}
		\State \textbf{Input:} A task identified by \texttt{ccai:taskName = ``View \& Update Competency Profiles''}.
		\State \textbf{Output:} Tuples (\texttt{?task}, \texttt{?process}, \texttt{?context}, \texttt{?resource}, \texttt{?role}, \texttt{?agent}).
		
		\begin{lstlisting}[style=sparqlstyle,belowskip=0pt]
PREFIX ccai: <http://gamaizer.ai/ccai#>
PREFIX prov: <http://www.w3.org/ns/prov#>			
SELECT DISTINCT ?task ?process ?context ?resource ?role ?agent
WHERE {
 ?task a ccai:Task ;
 ccai:taskName ``View & Update Competency Profiles'' .		
 # Process - Task
 OPTIONAL {
  { ?task ccai:partOfProcess ?process . }
 UNION
 { ?process a ccai:CollaborationProcess ;
  ccai:containsTask ?task . }
 }
 # Task - Context
 OPTIONAL { ?task ccai:hasContext ?context . }
 # Task <-> Resource
 OPTIONAL {
 { ?task ccai:includesResources ?resource . }
  UNION
  { ?resource a ccai:CollaborationResource ;
  ccai:usedForTask ?task . }
 }		
 # Task execution - Agent role
 OPTIONAL {
  ?agent ccai:executes ?task ;
  ccai:assignedRole ?role .
 }		
}\end{lstlisting}
	\end{algorithmic}
	\label{query:query_ex}
\end{algorithm}

\begin{table}[t]
	\centering
	\caption{Bindings returned for the task \textit{View \& Update Competency Profiles} in the context of \textit{ccai:Sprint1Context}.}
	\label{tab:query_view_update_results}
	\begin{tabular}{lll}
		\toprule
		?resource & ?role & ?agent \\
		\midrule
		ccai:CompetencyDB & ccai:DeveloperRole\_1 & ccai:HumanDeveloper\_Carol \\
		ccai:CompetencyDB & ccai:CodeAssistantRole\_1 & ccai:AICodeAssistant \\
		ccai:CompetencyDB & ccai:QAEngineerRole\_1 & ccai:HumanQA\_Lee \\
		ccai:UserAPI      & ccai:DeveloperRole\_1 & ccai:HumanDeveloper\_Carol \\
		ccai:UserAPI      & ccai:CodeAssistantRole\_1 & ccai:AICodeAssistant \\
		ccai:UserAPI      & ccai:QAEngineerRole\_1 & ccai:HumanQA\_Lee \\
		ccai:StyleGuide   & ccai:DeveloperRole\_1 & ccai:HumanDeveloper\_Carol \\
		ccai:StyleGuide   & ccai:CodeAssistantRole\_1 & ccai:AICodeAssistant \\
		ccai:StyleGuide   & ccai:QAEngineerRole\_1 & ccai:HumanQA\_Lee \\
		\bottomrule
	\end{tabular}
\end{table}

The PM initiates the ``\textit{View \& Update Competency Profiles}'' task, allocating resources to specific team members and verifying that the existing backlog properly captures all user requirements. The Technical Lead and UX/UI Designers develop wireframes and architectural guidelines for the user-facing interface, detailing the routes and APIs that will handle fetching and updating competencies. The developers collaborate with the AI Code Assistant by prompting it to generate skeleton code for the front-end and back-end endpoints. This collaborative iteration involves multiple round trips: developers specify constraints and desired functionalities, the AI generates draft code, and developers refine or merge the results into the main repository.

To maintain contextual consistency and an auditable trail, the team uses SPARQL queries to retrieve the operational context of the ongoing work as well as the resources and responsibility assignments associated with the task.
Query Box~\ref{query:query_ex} illustrates this retrieval for the \textit{View \& Update Competency Profiles} task.
For \textit{ccai:Sprint1Context}, the query returns three collaboration resources (\textit{ccai:CompetencyDB}, \textit{ccai:UserAPI}, \textit{ccai:StyleGuide}) and three collaborating roles with their associated agents
(\textit{ccai:DeveloperRole\_1} $\rightarrow$ \textit{ccai:HumanDeveloper\_Carol},
\textit{ccai:CodeAssistantRole\_1} $\rightarrow$ \textit{ccai:AICodeAssistant},
\textit{ccai:QAEngineerRole\_1} $\rightarrow$ \textit{ccai:HumanQA\_Lee}),
yielding 9 bindings in total (Table~\ref{tab:query_view_update_results}).
These bindings populate the contextual fields of the prompt template shown in Figure~\ref{fig_07}, including task scope, project context, supporting resources, role expectations, and expected output type. This provides the AI assistant with explicit, query-derived context rather than relying on free-form prompt elaboration.

For example, for the task ``View \& Update Competency Profiles'', the bindings returned by Query Box~\ref{query:query_ex} populate the prompt fields as follows: \textit{Task} = ``View \& Update Competency Profiles'', \textit{Context} = \textit{ccai:Sprint1Context}, \textit{Resources} = \{\textit{ccai:CompetencyDB}, \textit{ccai:UserAPI}, \textit{ccai:StyleGuide}\}, and \textit{Agents/Roles} = \{\textit{ccai:HumanDeveloper\_Carol}/\textit{DeveloperRole\_1}, \textit{ccai:AICodeAssistant}/\textit{CodeAssistantRole\_1}, \textit{ccai:HumanQA\_Lee}/\textit{QAEngineerRole\_1}\}. These values are then merged with the human instruction to produce the final prompt sent to the Generative AI assistant.

In this collaborative scenario, the Generative AI Agent serves as an on-demand code assistant and decision-support tool, refining outputs in real-time based on contextual information managed by the CCAI ontology. Using the structured prompt assembled from SPARQL-retrieved context, the Generative AI Agent serves as an on-demand code assistant and decision-support tool. The resulting prompt guides the AI's responses, ensuring that its suggestions align with the task's objectives and adhere to domain-specific constraints. As illustrated in Figure~\ref{fig_07}, the prompt template uses the contextual information retrieved via SPARQL (domain, timeline, team roles, resources, and requirements) to produce an AI Assistant prompt precisely tailored to the ``\textit{View \& Update Competency Profiles}'' feature. By embedding domain-specific context, team composition, relevant resources, and functional requirements into a single structured query, the prompt specifically requests implementation plans, code examples, testing strategies, and security considerations -- thus ensuring the AI's output directly addresses the project's immediate objectives while maintaining collaborative consistency across human and Generative AI stakeholders.

The generated output is then represented as an intermediate or final collaborative artifact and linked back to the originating task through provenance-aware relations. More specifically, the artifact can be connected to the task through \textit{prov:wasGeneratedBy} and attributed to the responsible human or AI contributor through \textit{prov:wasAttributedTo}. This closes the loop between task execution, SPARQL-based context retrieval, prompt construction, AI generation, and artifact traceability.

\begin{figure*}[h!]
	\centering	
	\includegraphics[width=5.1in]{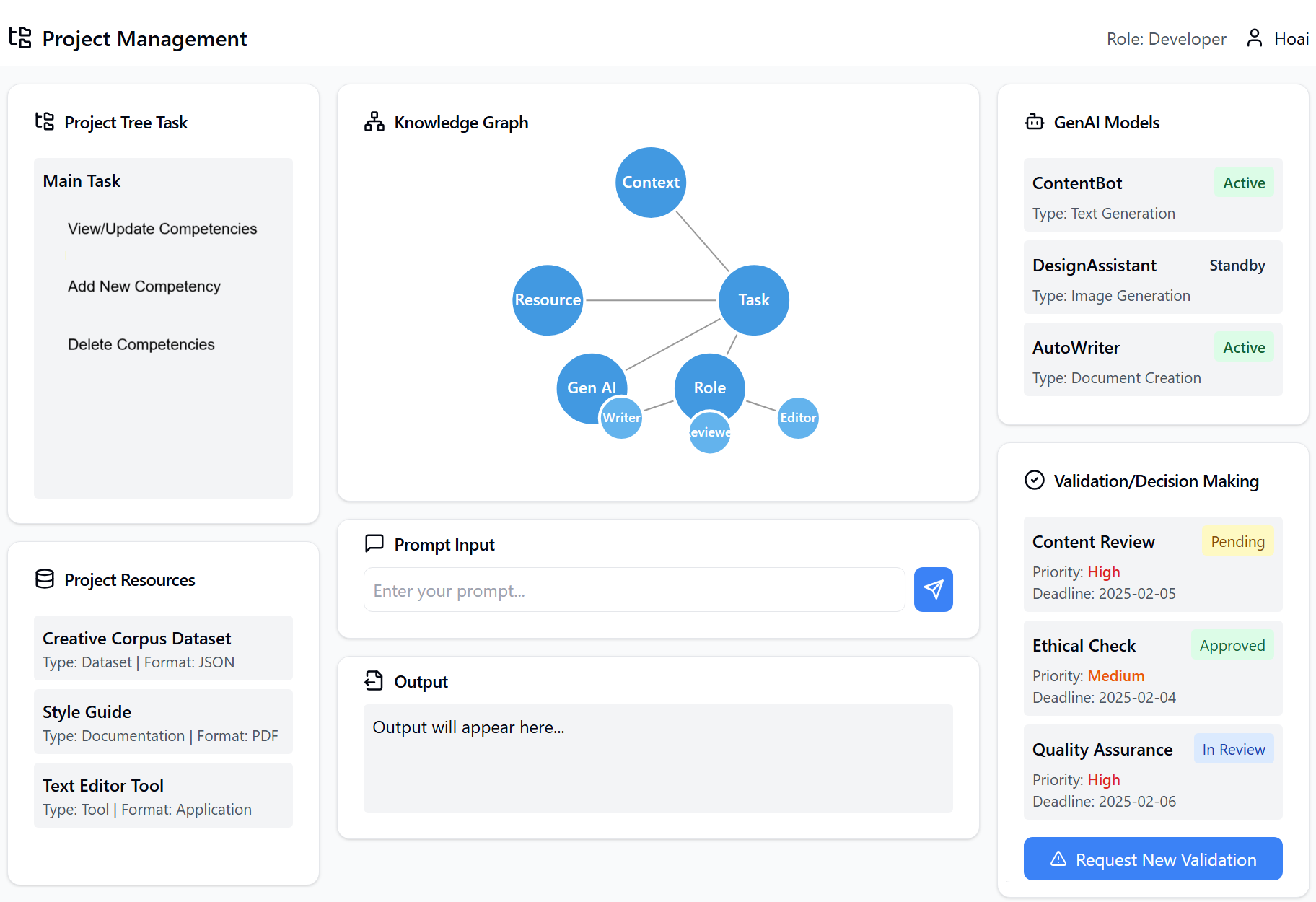}
	
	\caption{An integrated interface illustrating how human agents and Generative AI systems collaborate.}
	\label{fig_06}
\end{figure*}

In a broader view of the collaborative environment, the integrated interface, as illustrated in Figure~\ref{fig_06}, provides a project management setting where human collaborators -- such as the Developer ``Carol Hoai'' -- work alongside multiple Generative AI agents in an ontology-based visualisation for contextual information. The Project Tree Task panel outlines the system's main objectives (e.g., ``\textit{View/Update Competencies}'', ``\textit{Add New Competency}'', ``\textit{Delete Competencies}''), offering a structured roadmap for the team. Below that, the Project Resources panel highlights essential assets -- such as datasets and tool support -- that both humans and AI agents can reference during development. A Knowledge Graph visually maps the relationships among ``\textit{Collaboration Context}'', ``\textit{Collaboration Resource}'', ``\textit{Task}'', ``\textit{Generative AI}'', and ``\textit{Role}'', ensuring semantic consistency throughout the process. The Generative AI Models panel indicates which AI services (e.g., ContentBot, DesignAssistant, AutoWriter) are active, while the Validation/Decision Making panel shows ongoing reviews -- ranging from content checks to ethical validations and quality assurance -- each with distinct priorities and deadlines. Finally, a Prompt Input box at the bottom center lets users issue instructions to active AI agents, and an adjacent Output section displays the AI-generated responses in real time. This setup allows human collaborators to maintain strategic oversight and decision-making power, while AI tools provide on-demand content generation, testing support, or design assistance.

Figure ~\ref{fig_06} can also be interpreted as a managerial inspection interface. Beyond exposing collaboration context to contributors, it enables a project manager to analyze responsibility and process quality across AI-assisted tasks. Since generated outputs are connected to tasks, resources, prompts, and responsible agents, the manager can identify, for instance, which AI-generated code drafts most frequently required manual correction, which tasks systematically involved additional review cycles, or whether certain resource configurations were associated with weaker outputs. Such patterns are important because they transform provenance from a passive record into an operational monitoring mechanism. Rather than merely documenting that AI was used, the framework makes it possible to examine where AI support was effective, where it created rework, and where additional human oversight remained necessary.

The collaboration task for the ``\textit{View \& Update Competency Profiles}'' feature demonstrates how human stakeholders and Generative AI agents can work together to significantly accelerate development in coding, writing, and review processes. By leveraging contextual information from the ontology, the AI agents efficiently address routine or repetitive tasks, enabling human collaborators to focus on higher-level strategic objectives and overall system performance. 

\subsection{Key Findings from the Case Study}\label{sec:key-findings}

This section analyzes the application of the CCAI-based framework within our case study. We discuss key observations related to the research questions introduced in  \S\ref{sec:evaluation}, grounding our analysis in auditable traces (SPARQL results and ontology instances) and project artifacts (prompts, AI-generated tests, commits, and documentation).

In response to RQ1, the case study indicates that the ontology-based workflow helps reduce contextual ambiguity by externalizing task context, resources, and role assignments into queryable structures. Concretely, the SPARQL query in
Query Box~\ref{query:query_ex} returns a structured snapshot of the collaboration state for the
\textit{View \& Update Competency Profiles} task in \textit{ccai:Sprint1Context}: three required resources
(\textit{ccai:CompetencyDB}, \textit{ccai:UserAPI}, \textit{ccai:StyleGuide}) and three collaborating roles
(\textit{ccai:DeveloperRole\_1}, \textit{ccai:CodeAssistantRole\_1}, \textit{ccai:QAEngineerRole\_1}) mapped to their
responsible agents (\textit{ccai:HumanDeveloper\_Carol}, \textit{ccai:AICodeAssistant}, \textit{ccai:HumanQA\_Lee})
(Table~\ref{tab:query_view_update_results}). Rather than relying on free-form prompt elaboration, this produces a
fixed-size, machine-readable prompt context consisting of 9 bindings (3 resources $\times$ 3 role--agent assignments),
which can be injected directly into the prompt template (Figure~\ref{fig_07}). This shifts the interaction from iterative
context elicitation toward a single-step, context-complete request whose inputs are explicitly enumerated.

In response to RQ2, semantic annotations enable traceability by providing explicit provenance and responsibility links that
can be queried and reconstructed. In our case study, artifacts (e.g., AI draft outputs and project deliverables) are linked
to their responsible agents via provenance relations such as \textit{prov:wasAttributedTo} (Figure~\ref{fig_08}).
As a result, a reviewer can reconstruct a decision trail by traversing explicit graph links (artifact $\rightarrow$ agent,
task $\rightarrow$ agent, and associated context/resources), rather than relying on manual searches across documents and
repositories. This yields an auditable chain in which responsibility attribution is represented as query results rather
than narrative recollection.

In addressing RQ3, the primary limitation is the overhead required to keep the knowledge base synchronized with evolving
project resources and requirements. The case study includes a representative inconsistency scenario in which a resource
change was not reflected immediately in the ontology, causing the prompt context to contain
an outdated reference. This illustrates that the effectiveness of SPARQL-driven prompting depends on the timeliness of ABox
updates: when a required resource description is stale, the generated prompt context becomes partially incorrect and
human intervention is needed to correct the output. Consequently, ontology curation becomes an explicit operational task
in the workflow, with maintenance effort scaling with the rate of change of project artifacts (e.g., API specifications,
database schemas, and documentation).

These findings are bounded by a single, in-situ project and emphasize transparency, traceability, and shared understanding; we do not make efficiency claims. Credibility is supported through triangulation between artifact analysis and queryable traces, preserving auditable links from each claim to concrete evidence (queries, ontology instances, commits).
In the next section, we delve into a broader discussion of these collaborative achievements, their impact on project scalability, and the potential challenges arising from increased reliance on AI-driven workflows.
\section{Discussion}\label{sec:discussion}

Our findings underscore the effectiveness of integrating human expertise and Generative AI through a shared semantic framework. By structuring key concepts -- \textit{tasks}, \textit{roles}, \textit{resources}, and \textit{processes} -- the CCAI ontology helps all stakeholders operate within a coherent, transparent environment and maintain alignment between AI-generated outputs and evolving project requirements. In our case study, ontology-backed prompts (via SPARQL) and provenance links provided a unifying substrate that helped improve explanation clarity, make collaboration context more explicit, and support auditable decision trails across distributed teams.

The key takeaway is that semantic alignment is pivotal for effective Human-AI collaboration. In our study, the CCAI ontology operated as a boundary object \cite{star1989institutional}: a shared, inspectable representation that human developers can examine and AI agents can query. By anchoring work in shared contextual definitions (e.g., \textit{Competency}, \textit{Collaborator Role}, \textit{Collaboration Context}), the ontology helped reduce ambiguities that often affect multi-actor workflows. Practically, SPARQL-derived context guided human prompts and constrained AI outputs toward project goals, while relation-level justifications (e.g., \textit{ccai:usedForTask}, \textit{ccai:assignedRole}) enabled developers to verify recommendations before integration (cf. \S\ref{sec:key-findings}). In parallel, systematic provenance (e.g., \textit{prov:wasAttributedTo}) sustained traceability from user stories to generated tests and commits, fostering transparent auditing and accountable decision making. Together, these mechanisms stabilized collaboration by providing a semantic scaffold for interaction -- standing in clear contrast to unstructured, prompt-only practices where context drift and opaque rationales accumulate.


Based on our findings, we distill three design principles for practitioners and researchers creating future Human-AI collaborative systems. First, systems should make context explicit and inspectable rather than assuming an LLM can infer all task, role, and resource details. Exposing the underlying knowledge model -- ontology instances such as \textit{Task}, \textit{Collaborator Role}, and \textit{Collaboration Resource}, together with their relations -- allows practitioners to verify or correct the system's understanding before errors propagate. In our case study, prompts grounded in SPARQL-derived context helped limit drift and rework by aligning suggestions more closely with the project?s actual data model and APIs.

Second, accountability improves when provenance is integrated by default. Instead of relegating evidence to logs, significant AI outputs should be linked to their prompts, sources, and responsible agents through explicit relations (e.g., \textit{prov:wasAttributedTo}). Relation-level justifications surfaced in-line (e.g., via an explainer panel) helped reviewers reconstruct decision trails during audits and supported faster, more confident acceptance or rejection of AI suggestions.

Third, the ontology should serve as a boundary object for communication, not merely a backend schema. Lightweight query and visualization affordances enable both humans and Generative AI services to consult the same situational model, stabilizing terminology and expectations across roles. In practice, integrating SPARQL snippets, instance browsers, and rationale views into everyday interfaces allowed designers, developers, and reviewers to share up-to-date context and coordinate changes without ad-hoc translation layers.

Keeping the knowledge graph up to date becomes a shared operational responsibility rather than a one-time modeling effort. In realistic projects, this maintenance work is distributed across several actors. Developers may update links between tasks, resources, and implementation artifacts when APIs, schemas, or code dependencies evolve; project managers or technical leads may validate task assignments, responsibility relations, and workflow status; and domain experts may revise higher-level contextual or organizational constraints when requirements change. The effort required depends on the intended granularity of traceability: lightweight adoption may only require updates at major milestones, whereas finer-grained accountability requires more frequent synchronization with commits, generated artifacts, and documentation changes. At the same time, this burden can be partially reduced through semi-automatic trace capture from issue trackers, repositories, prompt logs, CI/CD pipelines, or generated outputs. In such a setting, human actors remain responsible for validating high-level semantic relations, while AI-assisted services can suggest candidate updates, detect stale references, and support the maintenance of domain-specific extensions of CCAI. This also opens the possibility of reusing CCAI itself to trace the evolution of these ontology extensions as part of the broader collaboration process.

This paper presents a single illustrative case study to demonstrate the feasibility and mechanics of our proposed framework. Our goal was to emphasize transparency, traceability, and shared understanding rather than to measure efficiency or productivity metrics. As such, we avoid claims about time-on-task or throughput.
This illustrative approach has inherent limitations. It demonstrates the potential of our framework but does not provide generalizable empirical evidence of its effects. Furthermore, the approach entails practical challenges. First, ontology design and maintenance demand ongoing effort in fast-evolving domains; controlling complexity is essential to sustain performance and developer adoption~\cite{osman2021ontology}. Second, although the ontology semantically grounds prompts and outputs, Generative AI models may still produce superficially plausible but domain-inadequate suggestions~\cite{teaming2022state}. Our feedback loops, including technical review and testing against real data, provide an important mitigation, but they do not eliminate this risk. More broadly, ontology-grounded prompting does not guarantee that the model will fully respect the requested structure, content, or constraints, since generation remains probabilistic. More conservative settings may improve consistency, but they can also limit diversity and creative exploration. This highlights an important trade-off in Human--Generative AI collaboration between tighter control and greater openness, depending on the nature of the task.

Future work will extend validation along two lines. Empirically, we plan a controlled comparison between teams using the ontology-backed assistant and teams using standard AI tools without an ontological backend, measuring quantitative outcomes such as task completion time, defect rates, code quality, and user-reported trust and workload. Methodologically, we will explore richer governance features and real-time constraints in the ontology (e.g., policy/risk rules, compliance checks, and performance monitors) to better support high-stakes settings~\cite{yun2024improving}. These steps will empirically validate the framework demonstrated in this paper, complementing our current analysis with quantitative evidence and strengthening the generality of the contribution.

\section{Conclusions and Perspectives}\label{sec5}
In this paper, we presented \emph{From Prompts to Context}, a Human--Generative AI Collaboration Framework enhanced by the CCAI ontology to support more dynamic, transparent, and traceable interactions between human stakeholders and Generative AI systems. By modelling roles, tasks, resources, constraints, and collaboration context within a shared ontology, we aimed to ensure that both human collaborators and Generative AI agents operate within a common semantic frame, reducing contextual ambiguities and aligning outputs with evolving project requirements. Our case study on implementing the ``\textit{View \& Update Competency Profiles}'' feature showed how ontology-backed prompts and explicit contextual links can create a persistent, queryable collaboration trail for artifacts and reviews, supporting clearer discussions and more accountable decision-making.
Despite these benefits, we encountered several challenges in maintaining and scaling the ontology, particularly in settings where terminology changes quickly or where integrations are technically complex. Keeping developers engaged with semantic modelling activities, while at the same time refining Generative AI behaviours to handle domain-specific subtleties, remains an open and sometimes delicate issue. Even when interactions are semantically grounded, model outputs may still look plausible on the surface yet fail to meet the deeper requirements of the domain, which reinforces the need for careful human review, testing, and validation alongside context-first practices.

Moving forward, we envision several avenues for further development: Advanced Ontology Governance, involving formal change management, version control, and policy attachment in rapidly evolving contexts; Model Interpretability and Validation, offering rationale surfaces and constraint checks that highlight ethical or domain-specific limits at the point of use; Scalability and Performance Optimization, exploring indexing and query techniques for large, distributed graphs; and Cross-Domain Integration, supporting interdisciplinary teams where bridging semantics across multiple fields can create greater synergy. By continually refining and expanding this approach, we believe that linking human expertise with Generative AI capabilities through the CCAI ontology will foster more adaptive, transparent, and ethically grounded collaborations in diverse sectors, from competency-based education management system to large-scale enterprise systems.

\backmatter





\bmhead{Acknowledgements}
We warmly thank the Ikigai consortium led by the association Games for Citizens, the company Gamaizer, as well as the FORTEIM project (winner of the AMI CMA France 2030 call for projects), for their support and collaboration. Their contributions have provided significant added value to the completion of this research.

\section*{Declarations}


\begin{itemize}
\item \textbf{Funding:} This research received no direct external funding. It benefited from support and collaboration from the Ikigai consortium led by the association Games for Citizens, Gamaizer, the Universit\'{e} de Technologie de Compi\`{e}gne (UTC), and the FORTEIM project (winner of the AMI CMA France 2030 call for projects).

\item \textbf{Conflict of interest/Competing interests:} The authors declare that they have no competing interests.

\item \textbf{Ethics approval and consent to participate:} Not applicable.
\item \textbf{Consent for publication:} Not applicable. No identifiable individual data are included in this manuscript.
\item \textbf{Data availability:} Not applicable.
\item \textbf{Materials availability:} The ontology and knowledge base created in this study are publicly available in the project's GitHub repository at \url{https://github.com/lengocluyen/ccai_ontology}.
\item \textbf{Code availability:} The code used to validate, and query the ontology/knowledge base is publicly available in the same GitHub repository at \url{https://github.com/lengocluyen/ccai_ontology}.
\item \textbf{Author contribution:} \textbf{\textit{Ng\d{o}c Luy\d{\^{e}}n \sur{L\^e}:}} Conceptualization; Methodology; Software; Ontology and knowledge base development; Data curation; Formal analysis; Investigation; Writing  - original draft; Visualization. \textbf{\textit{Marie-H\'{e}l\`{e}ne Abel:}} Conceptualization; Methodology; Supervision; Validation; Writing - review and editing. \textbf{\textit{Bertrand Laforge:}} Conceptualization; Supervision; Resources; Validation - review and editing.
\end{itemize}

\bibliography{sn-bibliography}

\end{document}